\documentclass[debug]{rmaa}

\usepackage{paralist}
\usepackage[english]{babel}
\usepackage{newtxtext,newtxmath}
\usepackage[T1]{fontenc}
\usepackage{ae,aecompl}
\usepackage{graphicx}	
\usepackage{amsmath}	
\usepackage{amssymb}
\usepackage{psfrag,color}
\usepackage[latin1]{inputenc}

\title{Can UVB variations reconcile simulated quasar absorption lines at high redshift?} 

\author{
  L. A. Garc\'ia,\altaffilmark{1,2,3}
  and E. V. Ryan-Weber \altaffilmark{2,3}}

\altaffiltext{1}{Grupo de Simulaci\'on, An\'alisis y Modelado, Vicerrector\'ia de Investigaci\'on, Universidad ECCI.}
\altaffiltext{2}{Centre for Astrophysics and Supercomputing, Swinburne University of Technology.}
\altaffiltext{3}{ARC Centre of Excellence for All-Sky Astrophysics (CAASTRO).}

\shortauthor{L. A. Garc\'ia et al.}
\shorttitle{UVB and HI self--shielding variations at $z =$ 6}

\fulladdresses{
\item  Luz \'Angela Garc\'ia: Universidad ECCI, Cra. 19 No. 49-20, Bogotá, Colombia, C\'odigo Postal 111311 (lgarciap@ecci.edu.co).
\item Emma V. Ryan-Weber: Swinburne University of Technology, Hawthorn, VIC 3122, Australia (eryanweber@swin.edu.au).}

\listofauthors{L. A. Garc\'ia \& E. V. Ryan-Weber}
\indexauthor{Garc\'ia, L. A.}
\indexauthor{Ryan-Weber, E. V.}

\abstract{In this work, we present new calculations of the observables associated with synthetic metal and HI absorption lines in the spectra of high redshift quasars, inspired by questions and limitations raised in work with a uniform Haardt-Madau 2012 ultraviolet background (UVB). We introduce variations at $z \sim$ 6 to the UVB and HI self--shielding and explore the sensitivity of the absorption features to modifications of the hardness of the UVB. We find that observed SiIV and low ionization states (e.g. CII, SiII, OI) are well represented by a soft UV ionizing field at $z =$ 6 but, this same prescription, fails to reproduce the statistical properties of the observed ion CIV absorber population. Therefore, we recommend a moderate reduction of the UVB at this redshift, an emissivity change between the UVB models that lies in between the Haardt-Madau 2012 emissivity J$_{\nu}$ and one with a dex below J$_{\nu} -$ 1 at 1 Ryd. On the other hand, variations in the HI self--shielding (SSh) prescription leave a non--negligible imprint in the calculated HI column density distribution function (CDDF) at $z =$ 4 and the comoving mass density of neutral Hydrogen (and the associated calculation with damped Lyman--$\alpha$ absorber systems) at 4 $< z <$ 6. We conclude that small variations in the UVB and HI SSh at $z \sim$ 6 play an important role in improving the estimation of metal ions and HI statistics at this redshift.}

\resumen{En este trabajo, presentamos nuevos resultados de los observables asociados a l\'ineas de absorci\'on sint\'eticas de HI y metales en el espectro de cu\'asares a alto redshift, inspirados por preguntas y limitaciones encontradas en un trabajo previo con el fondo ionizante uniforme (UVB) de Haard-Madau 2012. Introducimos variaciones del fondo ionizante y el modelo de \textit{self-shielding} del Hidr\'ogeno neutro y exploramos la sensibilidad de las l\'ineas de absorción al modificar la intensidad del UVB. Encontramos que los estados de baja ionizaci\'on (por ejemplo, CII, SiII, OI) y SiIV est\'an bien representadas por un UV ionizante m\'as suave, pero este mismo modelo falla en reproducir las propiedades estad\'isticas de la poblaci\'on observada de CIV. Por esto, recomendamos una reducci\'on moderada de la emisividad del fondo ionizante a este redshift, un order de magnitud por debajo. De otro lado, variaciones en el \textit{self-shielding} de HI dejan un marca en la funci\'on de distribuci\'on de las columnas de densidad en HI a $z =$ 4 y en la densidad de masa c\'osmica en el rango de 4 $< z <$ 6. Concluimos que peque\~nas variaciones en el fondo de UV y en el apantallamiento de Hidr\'ogeno neutro a $z \sim$ 6 juegan un papel importante en el mejoramiento de la estimaci\'on de las estad\'isticas de metales y HI a este redshift.}

\addkeyword{Cosmology: theory}
\addkeyword{Epoch of Reionization}
\addkeyword{Intergalactic medium}
\addkeyword{Methods: numerical}
\addkeyword{Quasars: metal absorption lines}

\begin{document}
\maketitle

\section{INTRODUCTION}
\label{sec:intro}
Unveiling the end of the Epoch of Reionization (EoR) and the sources that complete the budget of ionizing photons is currently a key topic in Astronomy. The phase transition of neutral Hydrogen (HI) into its ionized state (HII) was caused by the radiation released by the first stars \citep[POPIII,][]{abel2002,bromm2002,yoshida2003}, the second generation of stars \citep[POPII,][]{ciardi2005,mellema2006} and quasars \citep[with a black hole seed of 10$^6$ M$_{\odot}$,][]{dijkstra2004,hassan2017}. Other candidates are proposed, as miniquasars, with masses around 10$^{3-6}$M$_{\odot}$ \citep{mortlock2011,bolton2011,smith2017}, decaying or self--annihilating dark matter particles or decaying cosmic strings. Nonetheless, the latter objects seem to be unlikely to ionize the Universe by themselves.\newline

Understanding the EoR is intimately tied to the evolution of the ultraviolet background (UVB): the grand sum of all photons that have escaped from quasars and galaxies. Its spectral energy distribution is measured reasonably well at $z <$ 5 \citep{bolton2005} and modelled \citep{haardt2001}. \citet{haardt2012} used a cosmological 1D radiative transfer model that follows the propagation of H and He Lyman continuum radiation in a clumpy ionized intergalactic medium (IGM), and used mean free path and the hydrogen photoionization rate decrease with redshift \footnote{We refer to \citet{haardt2012} model as {\small{HM12}}.}. However, as $z \geq$ 6 is approached, the population of UV sources is not well determined \citep{haardt1999}. The uncertainty in estimating the UV photon emissivity from each type of object is caused by the lack of knowledge on the star formation rate, clumping factor and UV escape fraction \citep{cooke2014} at the redshift of interest, which are strongly model-dependent. Measuring Lyman series absorption or UV emissivity in a spectrum blueward of Ly$\alpha$ at 1216 $\AA$ is rendered almost impossible by the increasing density of matter and neutral Hydrogen fraction at redshifts greater than 5.5.\newline
On the other hand, the assumption of a uniform UV radiation field approximation breaks down close and during the EoR, when the interaction of the ionizing sources with the IGM requires a very accurate description \citep{lidz2006}. A real-time Reionization simulation should first ionize high density regions and fill some regions before than others, leading to a multiphase IGM with spatial fluctuations \citep{lidz2016}.\newline

Alternatively, intervening metal absorption lines detected in the spectra of high redshift quasars offer a completely different method to calibrate the UVB at high redshift. A growing number of absorption systems has been detected \citep{bosman2017,codoreanu2018,meyer2019,becker2019} and with the advent of bigger telescopes (e.g. GMT), the expectation is that the sample of absorption lines detected significantly increases. \newline
An increment in the sample of absorption lines (at least an order of magnitude with respect to the current observational sample) is possible with numerical simulations. A number of works have directed their efforts to describe the evolution of metal absorption lines in the intergalactic and circumgalactic medium (CGM). These simulations take into account different feedback prescriptions, photoionization modelling and variations in the UV ionizing background in the high redshift Universe  \citep{oppenheimer2006,oppenheimer2009,tescari2011,cen2011,finlator2013,pallottini2014,keating2014,finlator2015,rahmati2016,keating2016,garcia2017b,doughty2018, doughty2019}. The methods employed by each of these works, as well as the set up of the hydrodynamical simulations, show to have their advantages on the description of the IGM. \newline

However, findings in \citet{garcia2017b} with a uniform UV background \citep{haardt2012} show that the calculated column densities of the low ionization states (CII, SiII, OI) in the simulations struggle to match the values observed by \citet{becker2011}. The lack of spatial resolution on the scale of the low ionization absorbers evidences that further work needs to be done to reach a better description of the environment of these absorbers. Nonetheless, the uncertainties on the assumed UVB at high $z$ suggest that varying its normalization could alleviate the discrepancy between simulated results and the current observations. Works from \citet{finlator2015,finlator2016,doughty2018} also show that alternative models to the {\small{HM12}} UVB can reduce the gap between observables associated to metal lines calculated with their simulations and the observations. Their models account for simulated UVB with contribution from galaxies $+$ quasars and quasar--only. \newline

Triply ionized state of Silicon (SiIV) offers a unique avenue of investigation. Although it is classified as a high ionization state, its ionization potential energy is significantly lower than CIV, it does not necessarily lie in the same environment and exhibit the same physical conditions as CIV. Detections at high redshift of this ion have been made by \citet{songaila2001, songaila2005, dodorico2013}, \citet{boksenberg2015} and more recently by \citet{codoreanu2018}. The latter authors identify 7 systems across a redshift path of 16.4 over the redshift range 4.92 $< z <$ 6.13. They are $\sim$ 50$\%$ complete down to a column density of log N$_{\text{sys}}$ (cm$^{-2}$) of 12.50. This limiting column density allows them to study the identified SiIV population across the column density range $[$12.5, 14.0$]$ over the full redshift path of their survey. In addition, \citet{codoreanu2018} show that the fiducial configuration in \citet{garcia2017b} at $z =$ 5.6 is compatible with the SiIV observations.\newline

On the other hand, the self--shielding (SSh) of HI gas in very high density regions (above 10$^{17}$ cm$^{-2}$) is also a component that needs to be refined in the description, specifically when HI statistics are made. Current studies implement the self--shielding prescription proposed by \citet{rahmati2013a}, but unfortunately, this is only valid up to $z =$ 5. At redshifts when Reionization is concluding, this treatment is no longer valid. A new scheme has recently proposed by \citet{chardin2017b}, using radiative transfer calculations to find the best fitting parameters from the functional form of the photoionization rate $\Gamma_{\text{phot}}$ described in \citet{rahmati2013a}. Different SSh prescriptions could have a different outcome in the HI statistics. HI column densities are commonly classified in three regimes: Lyman--$\alpha$ forest (12 $<$ log N$_{\text{HI}}$ cm$^{-2} <$ 17.2), Lyman Limit systems (or LLS with 17.2 $<$ log N$_{\text{HI}}$ cm$^{-2} <$ 20.3) and Damped Lyman--$\alpha$ absorbers (DLAs) with N$_{\text{HI}} >$ 10$^{20.3}$ cm$^{-2}$. Works carried out by \citet{tescari2009, nagamine2004, pontzen2008, barnes2009, bird2014, rahmati2015, maio2015, crighton2015, garcia2017b} on DLAs have showed that DLAs are the main contributors to the cosmological mass density of HI. The HI self--shielding and molecular cooling prescriptions are important factors as these absorbers reside in low temperature and high density environments.\newline

This paper, in particular, builds on previous findings in \citet{garcia2017b}. The aim of this work is to explore and discuss to first approximation variations in the assumed UVB and the HI self--shielding prescriptions.\newline
The paper is presented as follows: Section~\ref{sec:sims} describes the simulations and the method used to post--process them. In section~\ref{sec:3} we explore two variations to the uniform {\small{HM12}} assumed in \citet{garcia2017b}. Additionally, we propose an alternative method to compare the current observations with the synthetic sample of metal absorbers, in contrast with previous works that compare two or more ions at once. Section~\ref{ssh} shows results for HI statistics when two different HI self--shielding prescriptions are implemented in post--process. Finally, Section~\ref{conclusions} summarizes the findings of this paper and explores the limitations encountered. 

\section{The numerical simulations and post-process}\label{sec:sims}

The results presented in this work are a follow-up to \citet{garcia2017b}, and are based on the simulations and the methodology presented in that paper. The suite of numerical simulations are with a customized version of GADGET-3 \citep{springel2005c}: P-GADGET3(XXL). The model was first tested in the context of the Angus project. In \citet{tescari2014} and \citet{katsianis2015,katsianis2016,katsianis2017}, the authors showed that their simulations are compatible with observations of the cosmic star formation rate density and the galaxy stellar mass function at 1 $< z <$ 7. The model takes into account the following physical processes: a multiphase star formation criterion; self-consistent stellar evolution and chemical enrichment modeling; Supernova (SN) momentum- and energy-driven galactic winds; AGN feedback, metal-line cooling; low-temperature cooling by molecules and metals \citep{maio2007,maio2015}. Moreover, the model is supported by: a parallel Friends-of-Friends FoF) algorithm to identify collapsed structures, and, a parallel  SUBFIND algorithm to identify substructures within FoF halos.\newline

The numerical model self-consistently follows the evolution of Hydrogen, Helium and 9 metal elements (C, Ca, O, N, Ne, Mg, S, Si and Fe) released from supernovae (SNIa and SNII) and low and intermediate mass stars. The chemical evolution scheme is based on the stochastic star formation model implemented in the simulations \citep{tornatore2007}. It accounts for the age of stars of different mass, such that the amount of metals released over time varies with the mass of the stars.\newline
The lifetime function from \citet{padovani1993} for stars with mass $m$ is adopted. The stellar yields quantify the amount of different metals which is released during the evolution of the stellar population, as follows: i) SNIa: \citet{thielemann2003}; ii) SNII (massive stars): \citet{woosley1995}; iii) Low and intermediate mass (AGB) stars: \citet{vandenhoek1997}.  As one of the main contributors to the Reionization of the Universe is the POPIII stars (a very massive and short-life population), they are best described by a \citet{chabrier2003} initial mass function (IMF).\newline

Galactic scale winds were introduced in GADGET simulations by \citet{springel2003} to regulate the star formation, spread metals from the galaxies and high-density regions to the IGM and shock-heated gas, and prevent the overcooling of gas.  The phenomenological model for energy-driven wind feedback is presented in \citet{springel2003}. It assumes that the mass-loss rate associated with the winds $\dot{M_w}$ is proportional to the star formation rate  $\dot{M_{*}}$, such that $\dot{M_w} = \eta \dot{M_{*}}$, with $\eta$ the wind mass loading factor that accounts for the efficiency of the wind. The kinetic energy of the wind is related to the energy input of the supernova. The velocity of the wind is given by the expression $v_w=2\sqrt{\frac{G M_{h}}{R_{200}}}= 2\times v_{\text{circ}}$. Due to the conservation of the wind energy, the velocity of the wind goes as the square of the inverse of the loading factor $\eta = 2 \left(\frac{600 \text{km/s}}{v_w}\right)^{2}$. \newline
However, \citet{puchwein2013} suggest that the  the mass carried by the wind is not necessarily proportional to the SFR of the galaxy.  In such case, it would be more natural to assume that there is a relation between the momentum flux (instead of the energy flux) of the wind and the SFR of the galaxy, thus $\eta$ is proportional to the inverse of the wind velocity $v_w$, such that $\eta = 2 \times \frac{600 \text{km/s}}{v_w}$.\newline

Chemical pollution caused by star formation contributes to the cooling of gas. Some metal line cooling efficiencies peak at $T \sim 10^4$ K \citep[mostly low ionization transitions,][]{gnat2012}. These transitions are privileged in metal poor high-density environments, as DLAs, where H is mostly neutral or in its molecular form. As discussed in \citet{maio2007}, molecular and low temperature metal cooling is particularly important when collapsed structures reach temperatures $T < 10^4$ K due to the formation of molecules. At this temperature, atomic cooling is not efficient and highly suppressed, yet, molecular H carries on cooling the gas.\newline

The assumed cosmology is a flat $\Lambda$CDM model with cosmological parameters from \citet{planck2015}: $\Omega_{0m}=$ 0.307, $\Omega_{0b}=$ 0.049, $\Omega_{\Lambda}=$ 0.693 and $H_0=$ 67.74 km s$^{-1}$Mpc$^{-1}$ (or $h =$ 0.6774). 
The simulations considered in the paper are described in Table~\ref{table_sims}, with comoving box size and softening of 18 Mpc/$h$ and 1.5 kpc/$h$, respectively. All the simulations have the same initial number of gas and DM particles ($2\times512^3$), with mass of gas and dark matter particles of $M_{\text{gas}} =$ 5.86 $\times$ 10$^{5}M_{\odot}$/$h$ and  $M_{\text{DM}} =$ 3.12 $\times$ 10$^{6}M_{\odot}$/$h$.  Moreover, we include molecular and low-temperature metal cooling \citep{maio2007, maio2015} in our simulations. The fiducial run is labelled Ch 18 512 MDW.\newline

\begin{table}[t!]
\caption{\footnotesize{Summary of the simulations used in this work.}}
\footnotesize{Column 1: run name. Column 2: box size. Column 3: Plummer-equivalent comoving gravitational softening length. Column 4: feedback model. Column 5: inclusion of low-temperature metal and molecular cooling \citep{maio2007, maio2015}. The first run, Ch 18 512 MDW, is the fiducial model. The second one in the list, Ch 18 512 MDW mol, has exactly the same configuration as the reference run, but includes low-T metal and molecular cooling.}
\label{table_sims}
\resizebox{0.95\textwidth}{!}{%
\centering
\begin{tabular}{lcccc} 
  \hline
Simulation &  Box size & Comoving softening &  Model for &  low-T metal \& \\
 &  (cMpc/$h$) & (ckpc/$h$) & SN--driven winds & molecular cooling\\ \hline
\textbf{Ch 18 512 MDW}  & \textbf{18} & \textbf{1.5} & \textbf{Momentum--driven} &    \\ 
Ch 18 512 MDW mol & 18 & 1.5 & Momentum--driven &  \checkmark\\ 
Ch 18 512 EDW & 18 & 1.5 & Energy--driven &   \\ 
\hline
\end{tabular}%
}
\end{table}

The numerical simulations are post--processed to recreate the observations of high redshift quasars and recover synthetic spectra for each ion along each sightline. The pipeline derived for this purpose relies on the physical conditions of the gas (reproduced with the hydrodynamical simulations), on top of a uniform field radiation that accounts for the cosmic microwave background (CMB) and the ultraviolet/X-ray background from quasars and galaxies {\small{HM12}} \citep{haardt2012}. The photoionization modeling for the metal transitions is computed with {\small{CLOUDY}} v8.1 \citep{ferland2013} for optically thin gas. In addition, a HI self-shielding prescription is imposed to the simulations to account for neutral Hydrogen inside high-density regions where gas is optically thick. We choose a thousand random lines of sight along to the three perpendicular directions inside the box, and in each one of these sightlines, we calculate a simulated spectrum, containing relevant physical information: HI flux/optical depth, density and temperature of the gas, and the number density of all the ions considered in the analysis, among other quantities. The box size $\Delta v$ at a given redshift is translated to the equivalent redshift path through the relationship $\Delta z = (1 + z) \frac{\Delta v}{c}$. Once the synthetic spectra are computed, they are convolved with Gaussian noise profiles with full width at half maximum $\text{FWHM} =$ 7 \mbox{km s$^{-1}$}. Finally, the individual absorption line features are fitted with Voigt profiles with the code {\small{VPFIT}} v.10.2 \citep{vpfit}. We focus our attention on the ionic transitions shown in Table~\ref{table_ions}.\\

\begin{table}
\caption{\footnotesize{List of the ion lines included in this work.}}
\footnotesize{The first column contains the ions, the second one the rest-frame wavelength $\lambda$ of the transition with the highest oscillator strength. The third column, the oscillator strength $f$ of each absorption line, the fourth one shows the ionization energy $E$ associated to each state, and the fifth column, the metal abundance log $Z$ (in solar units), taken from \citet{asplund2009}.}
\label{table_ions}
\vspace{0.2cm}
\begin{center}
\begin{tabular}{ccccc} 
\hline
Ion $i$ &  $\lambda$ (\AA) & $f$ & $E_{i \rightarrow i+1}$ (eV) & log $Z$ ($Z_{\odot}$)\\  \hline
HI & 1215.67  & 0.4164 & 13.6 & 0\\ 
CII & 1334.53  & 0.1278 &  24.38 & -3.57\\ 
CIV & 1548.21  & 0.1899 &  64.49 & -3.57\\ 
SiII & 1526.71  & 0.1330 &  16.35 & -4.49\\ 
SiIV & 1393.76  & 0.513 &  45.14 & -4.49\\ 
OI  & 1302.17 & 0.0480 & 13.62 & -3.31\\ 
\hline 
\end{tabular} \\
\end{center}
\vspace{0.2cm}
\footnotesize{Note: The energy $E$ shown in the fourth column is the energy required to reach the next ionization state $i+1$ from the state $i$.}\\
\end{table}

\section{Variations of the UVB in post-process}
\label{sec:3}

Investigations carried out with a uniform UV background \citep{haardt2012} in \citet{garcia2017b} showed that the calculated column densities of the low ionization states (CII, SiII, OI) and their corresponding observables (comoving mass density, column density distribution function, etc.) have some room for improvement when compared with observations. There is general agreement that current simulations do not have enough resolution on the scale of the absorbers with low ionization energies. However, uncertainties on the high $z$ UVB suggest that varying its normalization is a first step towards a better agreement with the observations at high redshift. \newline

Here we explore the sensitivity of the results presented in \citet{garcia2017b} to the presence of different ultraviolet/X-ray ionizing backgrounds by modifying the normalization factor at 1 Ryd of the uniform {\small{HM12}} UVB in post-process. Preliminary work allows us to conclude that decreasing the UVB intensity at $z =$ 6 is equivalent to imposing an ionizing background at earlier times that redshift 6. Therefore, low ionization ions would prevail in the early stages of the Epoch of Reionization. Hence, reductions of more than one order of magnitude in the UVB are very aggressive for high ionization species and simultaneously, lead to an overproduction of the low ionization ones. Consequently, a variation of 1 dex below the fiducial emissivity in HM12 is conservative but offers a non-negligible imprint on the calculated ions.\newline

\begin{figure*}[!t]
\centering
  \includegraphics[width=0.65\linewidth]{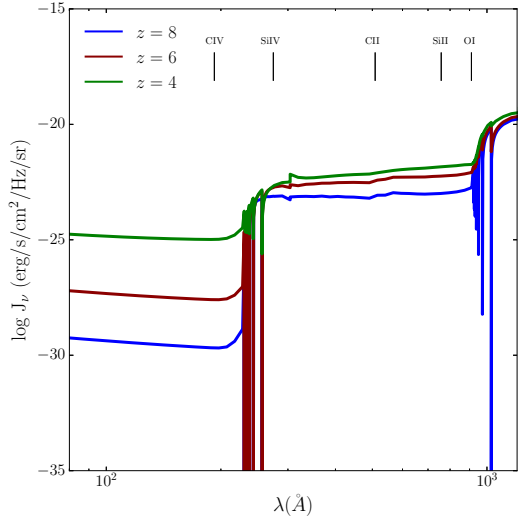}
  \caption{UV emissivity for the uniform {\small{HM12}} background at three different redshifts: $z =$ 8, 6 and 4 (blue, dark red and green, respectively) compared with the wavelength of the radiation, in the wavelength range where the ions transitions occur.}
\label{fig:hm12uniE}
\end{figure*}

Variations of the UVB spectrum (quasars+galaxies model) from the original {\small{HM12}} (see Figure~\ref{fig:hm12uniE}) require modified input files to run new {\small{CLOUDY}} tables.
\newline

The procedure followed here is explained in \ref{sec:sims} and it has been tuned and applied in \citet{garcia2017b,garcia2017a}, with some small adjustments: the normalization parameter at 1 Ryd  is reduced by an order of magnitude below the fiducial value in {\small{HM12}} at all redshifts. This leads to a softer UVB. Hereafter, the test is refered as \textbf{log J$_{\nu} -$ 1} (see section~\ref{section_uvb_variation}).

\subsection{Change in the normalization of HM12}
\label{section_uvb_variation}
It is a reasonable expectation that the presence of a softer UVB input than the uniform {\small{HM12}} in the photoionization model would favour low ionization states and more neutral states would show large incidence rates. In order to test this hypothesis, the UV emissivity J$_{\nu}$ at 1 Ryd  is reduced by one dex compared with the value defined by the uniform {\small{HM12}} at all redshifts. This leads to a softer UVB. \\

In order to avoid introducing more variables to this test, the box size has been fixed to runs with 18 Mpc/$h$. The simulations used to recover the observables are Ch 18 512 MDW, Ch 18 512 MDW mol and Ch 18 512 EDW. The convergence and resolution tests are not included in this document, but they can be found in \citet{garcia2017b}. \newline

The first comparison with the observations explored here is the CIV column density distribution function (CDDF), defined in equation~\ref{cddf}, as follows: \newline
\begin{equation}\label{cddf}
f(\text{N},X)=\frac{ n_{\text{sys}}(\text{N}, \text{N}+\Delta \text{N})}{n_{\text{lov}} \Delta X}.
\end{equation}
\noindent Here, $n_{\text{lov}}$ is the number of lines of view considered. The absorption path $\Delta X= \frac{H_0}{H(z)}(1+z)^2 \Delta z$ relates the Hubble parameter at a given redshift $z$ with the correspondent redshift path $\Delta z = (1+z)\frac{\Delta v}{c}$. The term $\Delta v$ is the box size in km s$^{-1}$.\newline

\begin{figure}[h!]
\centering
	\includegraphics[width=0.65\linewidth]{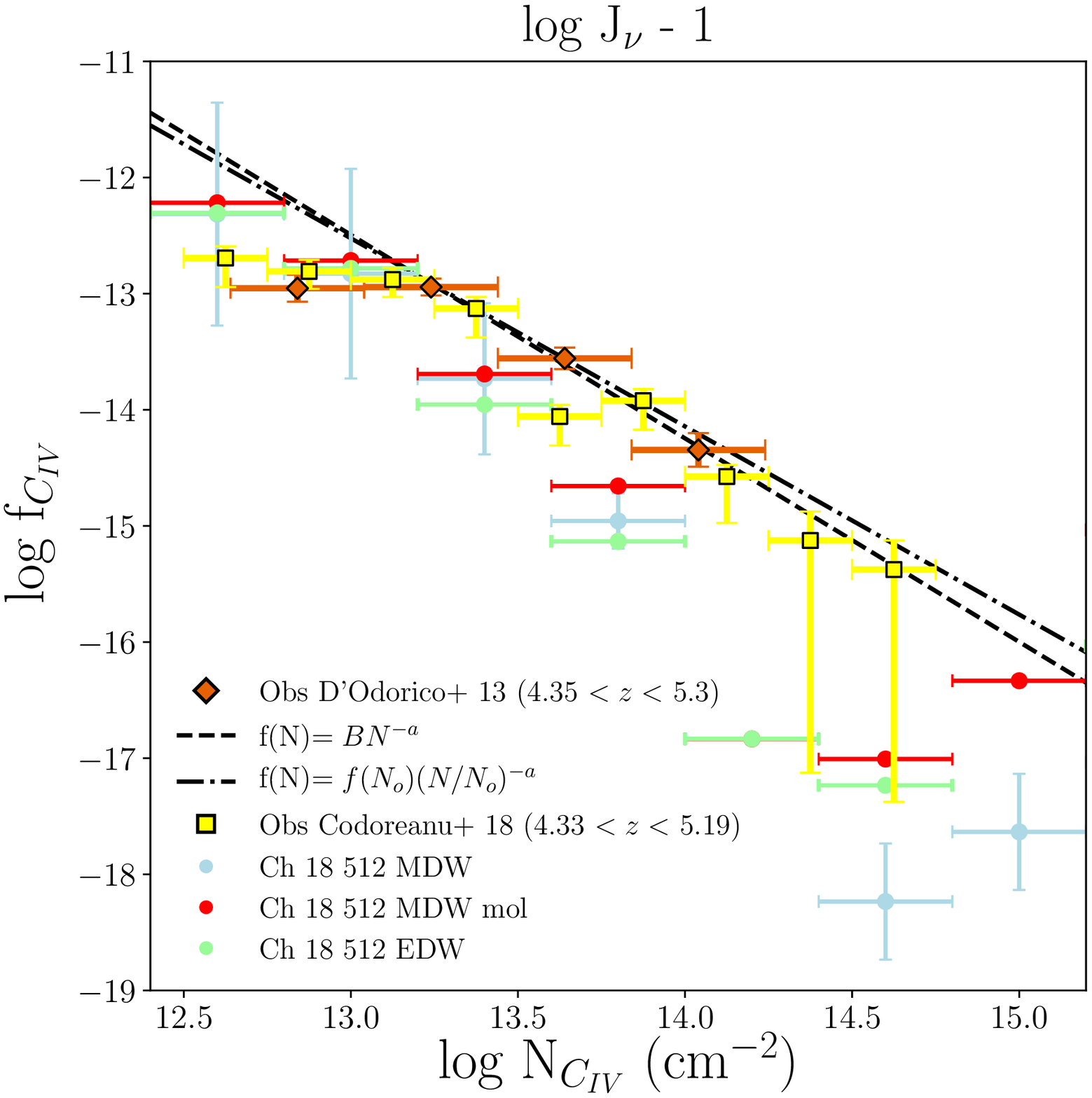}
       \caption{CIV column density distribution function at $z =$ 4.8 and comparison with observational data by \citet{dodorico2013} in orange diamonds and \citet{codoreanu2018} in yellow squares. The black dashed line represents the fitting function $f($N$) = B$N$^{-\alpha}$ with $B = $ 10.29 $\pm$ 1.72 and $\alpha =$ 1.75 $\pm$ 0.13 and the dotted--dashed line $f($N$) = f($N$_0)($N$/$N$_0)^{-\alpha}$ with $f($N$_0) = $ 13.56 and $\alpha =$ 1.62 $\pm$ 0.2, from \citet{dodorico2013} work. The error bars are the Poissonian errors for the reference run and are a good representation of the errors in the other models. The left panel shows results with the uniform {\small{HM12}} and the right panel the results with the test \textbf{log J$_{\nu} -$ 1}, for the simulations Ch 18 512 MDW, Ch 18 512 MDW mol and Ch 18 512 EDW. In all cases, but in particular in simulations without molecular cooling implemented, the number of CIV absorbers in the \textbf{log J$_{\nu} -$ 1} case is under-represented in the range of column densities considered.}
\label{fig:c4_cddf_48_5}       
\end{figure}

\begin{figure}[h!]
\centering
	\includegraphics[width=0.65\linewidth]{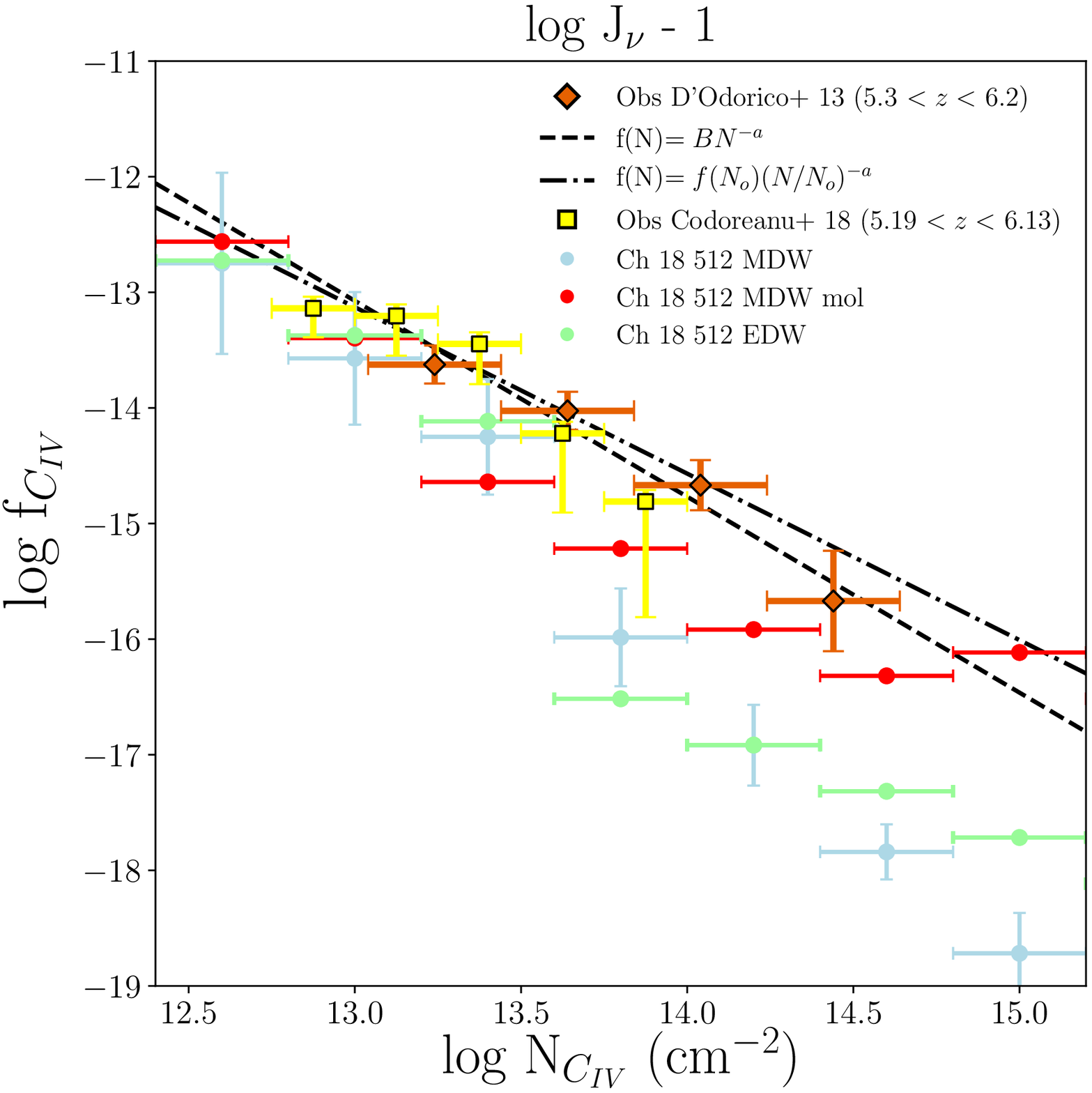}
	 \caption{CIV column density distribution function at $z =$ 5.6 and comparison with observational data by \citet{dodorico2013} in orange diamonds and \citet{codoreanu2018} in yellow squares. The black dashed line represents the fitting function $f($N$) = B$N$^{-\alpha}$ with $B = $ 8.96 $\pm$ 3.31 and $\alpha =$ 1.69 $\pm$ 0.24 and the dotted--dashed line $f($N$) = f($N$_0)($N$/$N$_0)^{-\alpha}$ with $f($N$_0) = $ 14.02 and $\alpha =$ 1.44 $\pm$ 0.3, from \citet{dodorico2013} work. The blue error bars are the Poissonian errors for the reference run and are a good representation of the errors in the other models. The left panel shows results with the uniform {\small{HM12}} and the right panel the results with the test \textbf{log J$_{\nu} -$ 1}, for the simulations Ch 18 512 MDW, Ch 18 512 MDW mol and Ch 18 512 EDW.}
	 \label{fig:c4_cddf_56_5}
\end{figure}

\begin{figure}[t!]
\centering
	\includegraphics[width=0.48\linewidth]{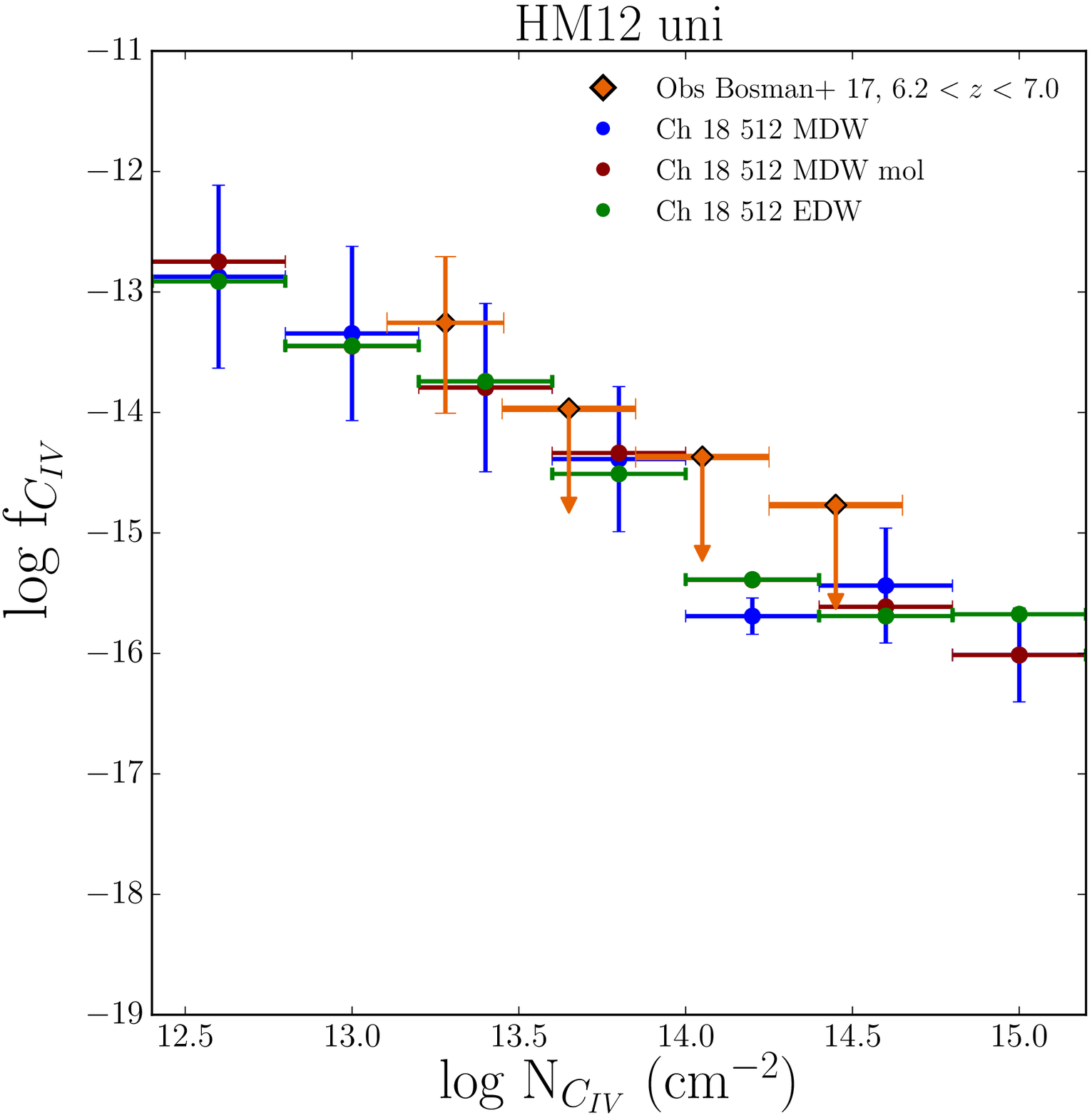}
	\includegraphics[width=0.45\linewidth]{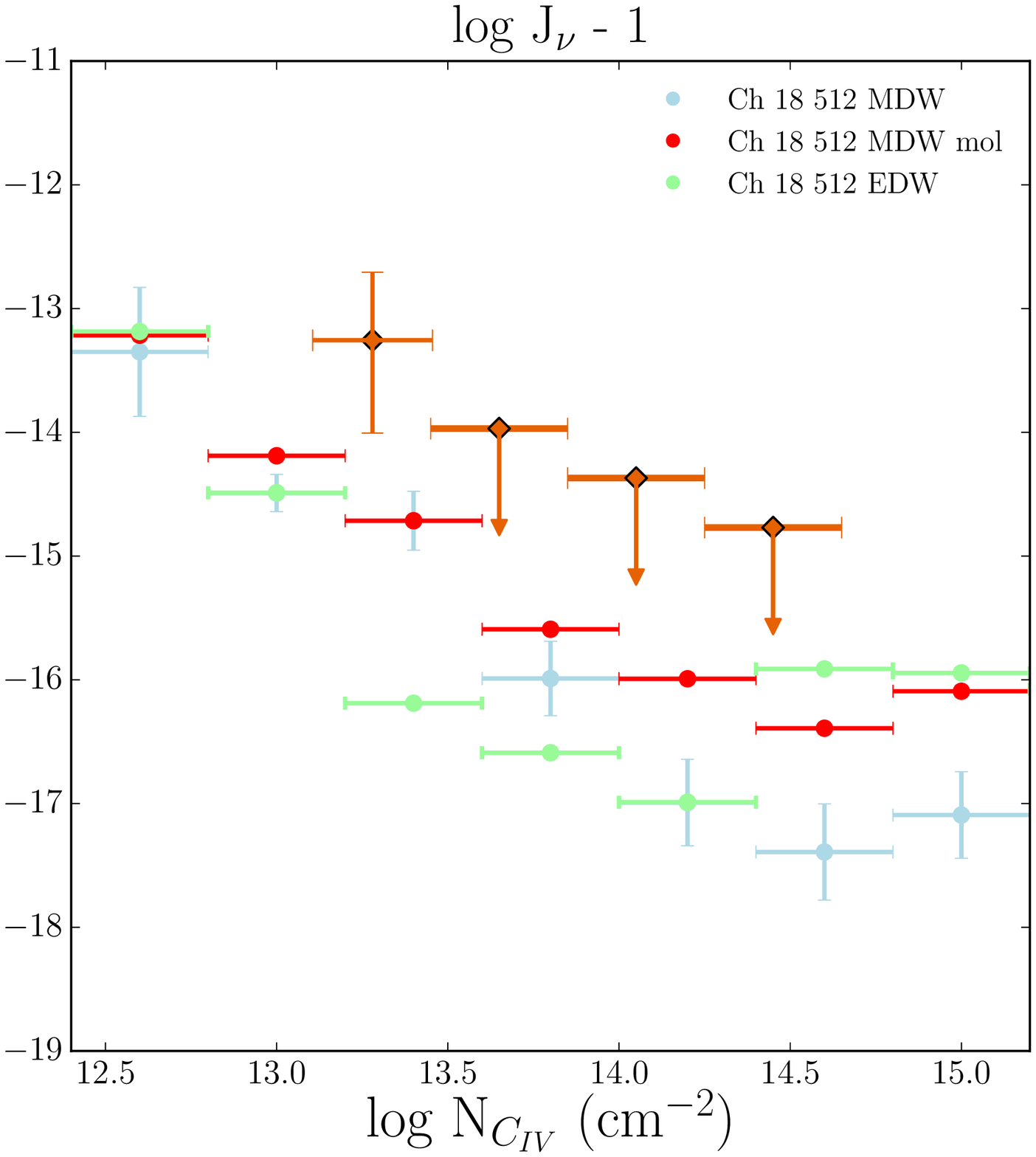}
\caption{CIV column density distribution function at $z =$ 6.4 and comparison with observational data by \citet{bosman2017} in orange diamonds. The error bars are the Poissonian errors for the reference run and are a good representation of the errors in the other models. The left panel shows results with the uniform {\small{HM12}} and the right panel the results with the test \textbf{log J$_{\nu} -$ 1}, both for the simulations Ch 18 512 MDW, Ch 18 512 MDW mol and Ch 18 512 EDW. }
\label{fig:c4_cddf_64_5}
\end{figure}

At $z =$ 4.8 and 5.6, the CCDFs are compared with observations from \citet{dodorico2013} and \citet{codoreanu2018} in Figures~\ref{fig:c4_cddf_48_5} and \ref{fig:c4_cddf_56_5}, respectively. At $z =$ 6.4, the simulated values of the CDDF are compared with upper limits from \citet{bosman2017} in Figure~\ref{fig:c4_cddf_64_5}. \newline

The key feature of Figures~\ref{fig:c4_cddf_48_5}, \ref{fig:c4_cddf_56_5}, \ref{fig:c4_cddf_64_5} is that all of them show a notable underproduction of the CIV absorbers at all redshifts when the emissivity is decreased. The calculated CDDFs are always below the observed values, and for high column densities, there is a clear departure from the fitting functions provided by \citet{dodorico2013}. Nonetheless, the synthetic CDDFs have a closer match with the observational values from \citet{codoreanu2018}, in particular at high column densities. It is worth noting that \citet{bosman2017} data are just upper limits for the CIV--CDDF at 6.2 $< z <$ 7.0, nevertheless the computed values in this test are significantly underrepresented. \newline

There is a change in the number of CIV absorbers when the UVB emissivity is lower than the original HM12. This result gives a hint in regards to the number of the CIV absorbers: the test \textbf{log J$_{\nu} -$ 1} is too aggressive with high ionization states of C lying in the IGM.\\


Next, we compute the cosmological mass density $\Omega_{\text{CIV}}$. The cosmological mass density of an ion is defined as:
\begin{equation}\label{omegaion}
\Omega_{\text{ion}}(z)=\frac{H_0 m_{\text{ion}}}{c \rho_{\text{crit}}}\frac{\sum \text{N}(\text{ion},z)}{n_{\text{lov}}\Delta X},
\end{equation}
\noindent with $m_{\text{ion}}$ the mass of the ionic species, $n_{\text{lov}}$ is the number of lines of view, $\rho_{\text{crit}}$ is the critical density today and $\Delta X$ is expressed above.\newline

\begin{figure}[h!]
\centering
	\includegraphics[width=0.48\linewidth]{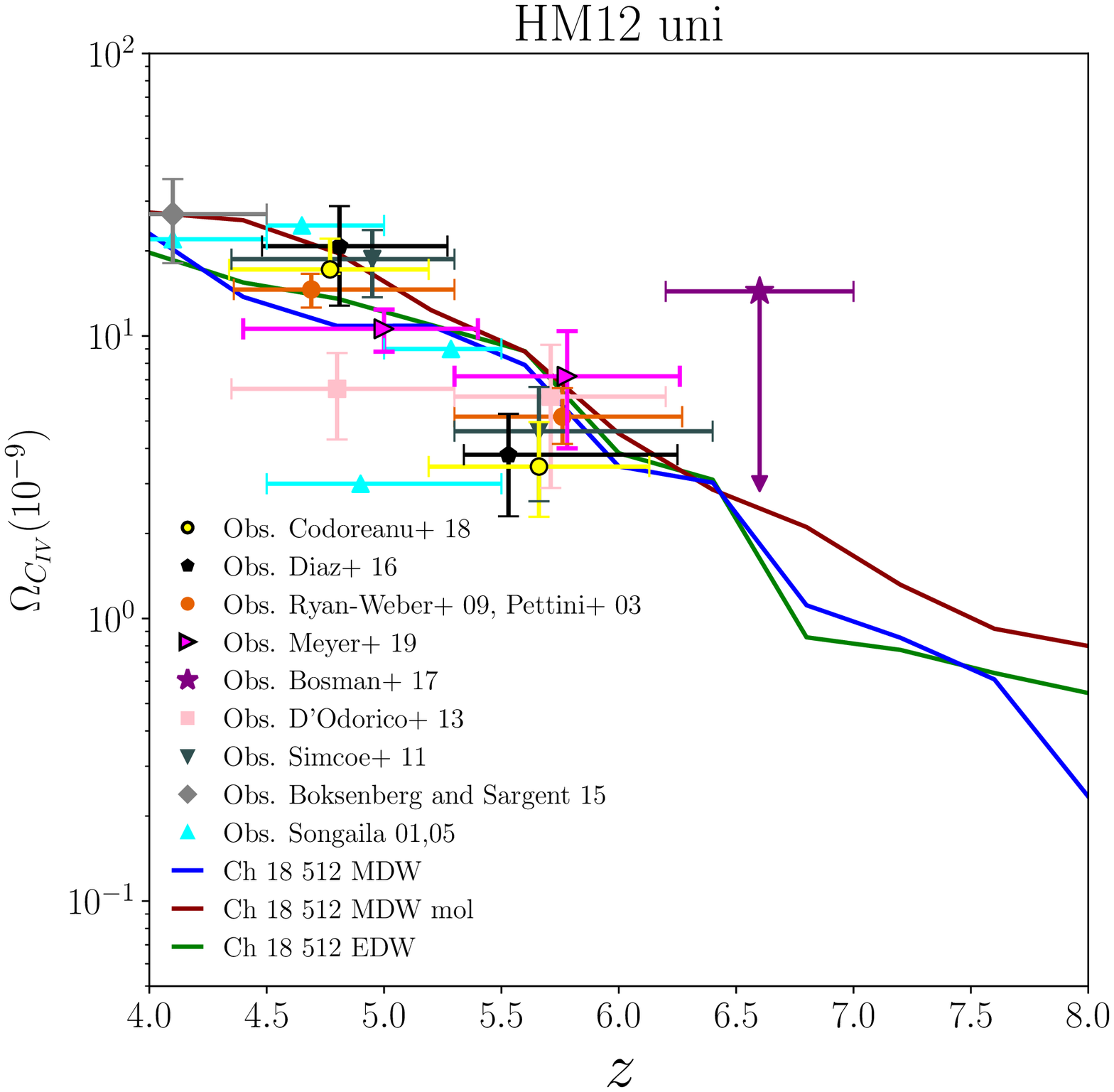}
	\includegraphics[width=0.45\linewidth]{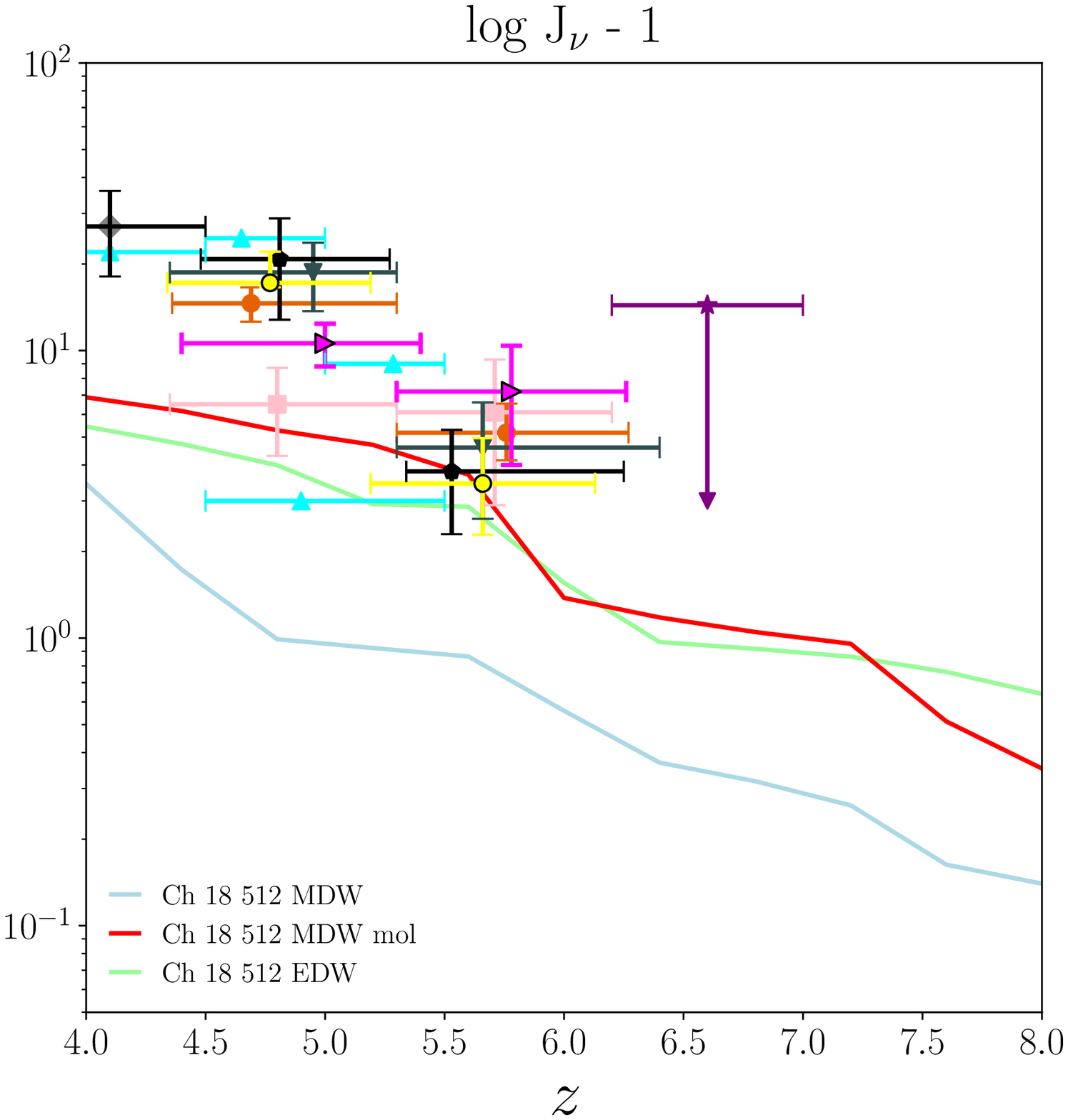}
        \caption{CIV cosmological mass density at 4 $< z <$ 8 for 13.8 $<$ $\log$ N$_{\text{CIV}}$(cm$^{-2}$) $<$ 15.0. Comparison between the simulated data and observations by \citet{pettini2003} and \citet{ryanweber2009} in orange circles, \citet{codoreanu2018} in yellow circles, \citet{songaila2001,songaila2005} in cyan triangles, \citet{meyer2019} in magenta triangles, \citet{simcoe2011} in dark green inverted triangles, \citet{dodorico2013} in pink squares, \citet{boksenberg2015} in grey diamond, upper limits from \citet{bosman2017} in purple star and \citet{diaz2016} in black pentagons. Pettini, Ryan-Weber, Codoreanu and D\'iaz measurements are converted to the Planck cosmology, while for the others this recalibration was not possible due to missing details of the precise pathlength probed. On the left panel the results with the uniform {\small{HM12}} are presented. The right panel shows $\Omega_{\text{CIV}}$ in the framework of the test \textbf{log J$_{\nu} -$ 1}. In both cases, the simulations used are Ch 18 512 MDW, Ch 18 512 MDW mol and Ch 18 512 EDW. As a consequence of the low number of absorbers in this column density range, when the UVB normalization is varied, $\Omega_{\text{CIV}}$ is at least an order of magnitude lower than the case with the original UVB.}
\label{fig:omega_nor}
\end{figure}

\begin{figure}[h!]
\centering
	\includegraphics[width=0.48\linewidth]{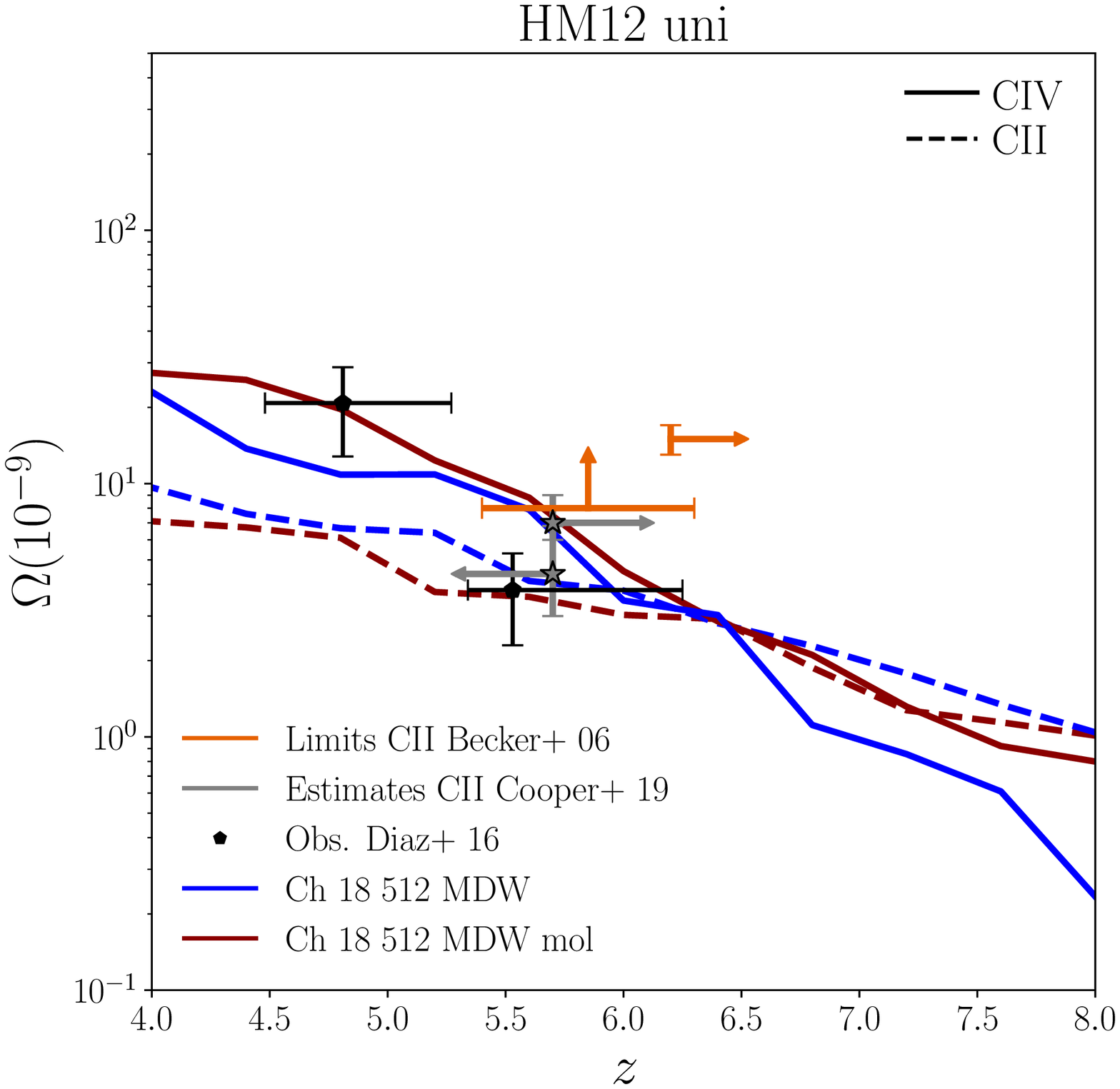}
	\includegraphics[width=0.48\linewidth]{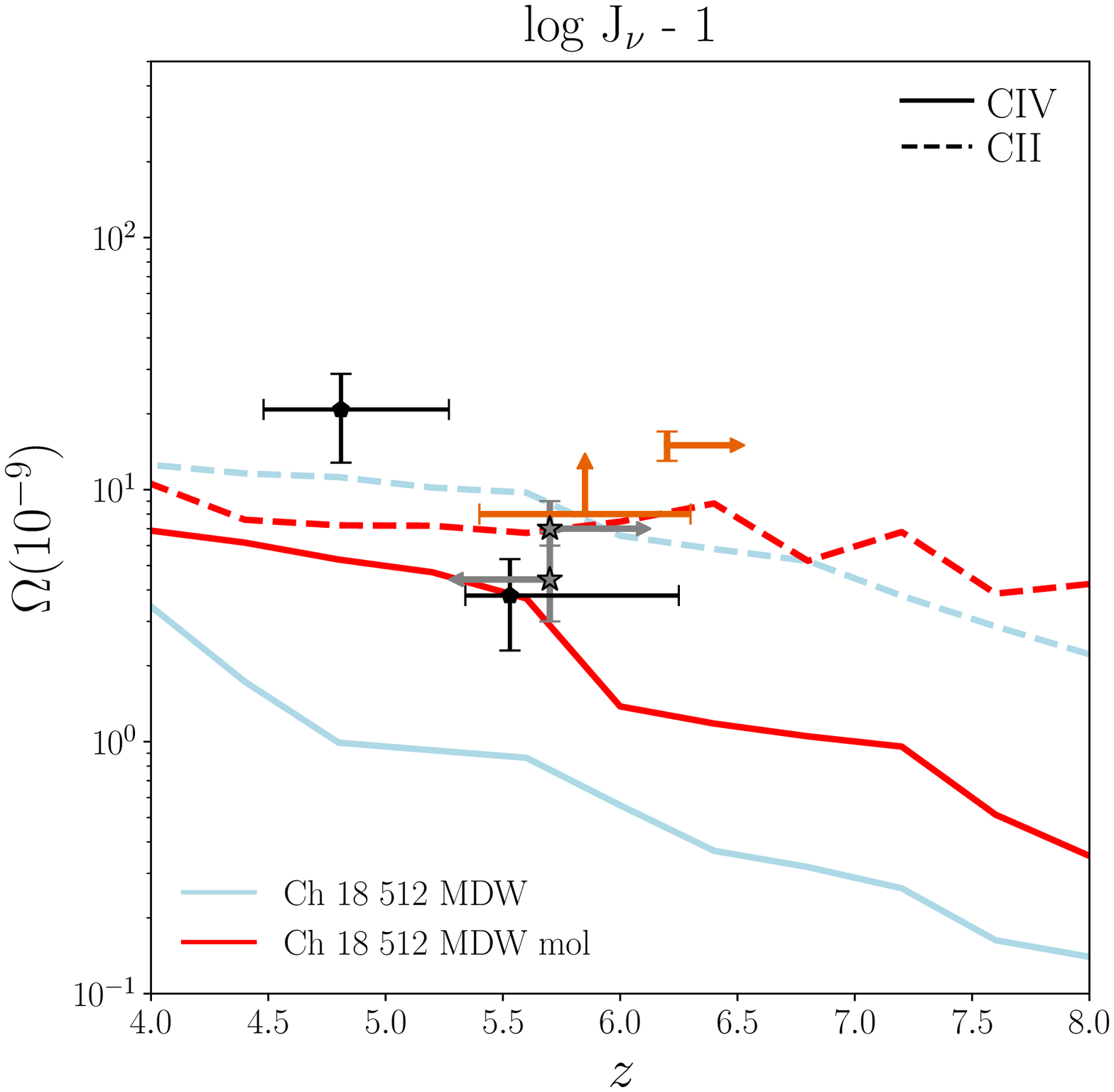}
        \caption{Evolution of the CII and CIV cosmological mass density when the normalization of the UVB is varied at 1 Ryd (comparison of molecular cooling content). On the left panel the results with the uniform {\small{HM12}} are presented. The right panel shows results of the test \textbf{log J$_{\nu} -$ 1}. In both cases, the runs used are Ch 18 512 MDW and Ch 18 512 MDW mol. The solid lines mark the evolution of $\Omega_{\text{CIV}}$ for 13.8 $<$ $\log$ N$_{\text{CIV}}$(cm$^{-2}$) $<$ 15.0, and the dashed lines $\Omega_{\text{CII}}$ in the range 13.0 $<$ $\log$ N$_{\text{CII}}$(cm$^{-2}$) $<$ 15.0. The orange points with errors represent the observational lower limits for $\Omega_{\text{CII}}$ from \citet{becker2006} and the grey arrows the corresponding estimates made by \citet{cooper2019}. The latter estimates have been done for $z >$ 5.7 and $z <$ 5.7, in right and left arrows, respectively. Although in the case with softer UVB there is not a crossover of CII and CIV (due to the low number of CIV absorbers), the mass density of CII matches the limits from \citet{becker2006} and \citet{cooper2019} in both simulations and CIV matches the observational detection at $z =$ 5.7 by \citet{diaz2016} in the molecular cooling run.}
        \label{fig:omega_mol_both}
\end{figure}

\begin{figure}[h!]
\centering
	\includegraphics[width=0.48\linewidth]{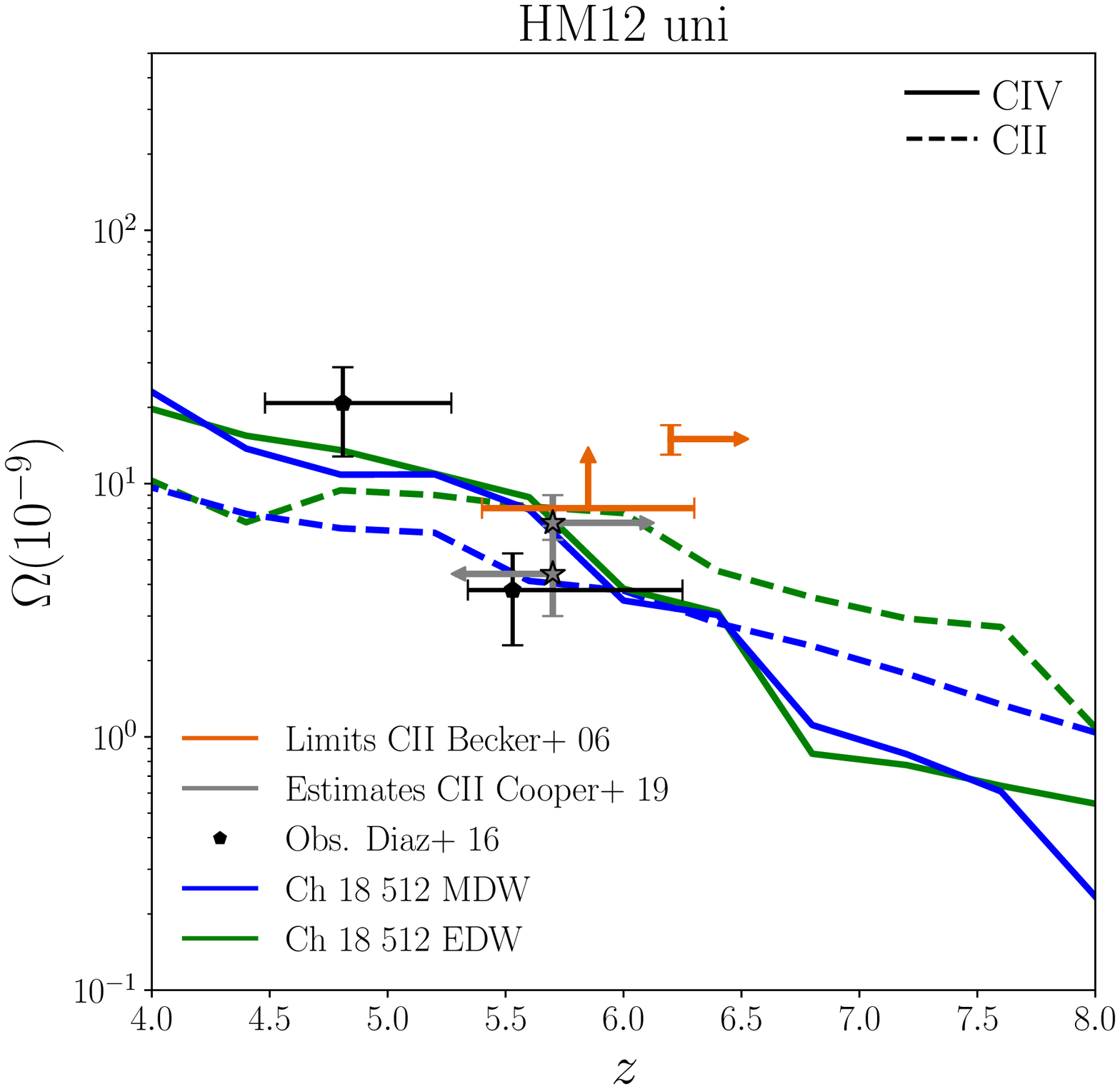}
	\includegraphics[width=0.48\linewidth]{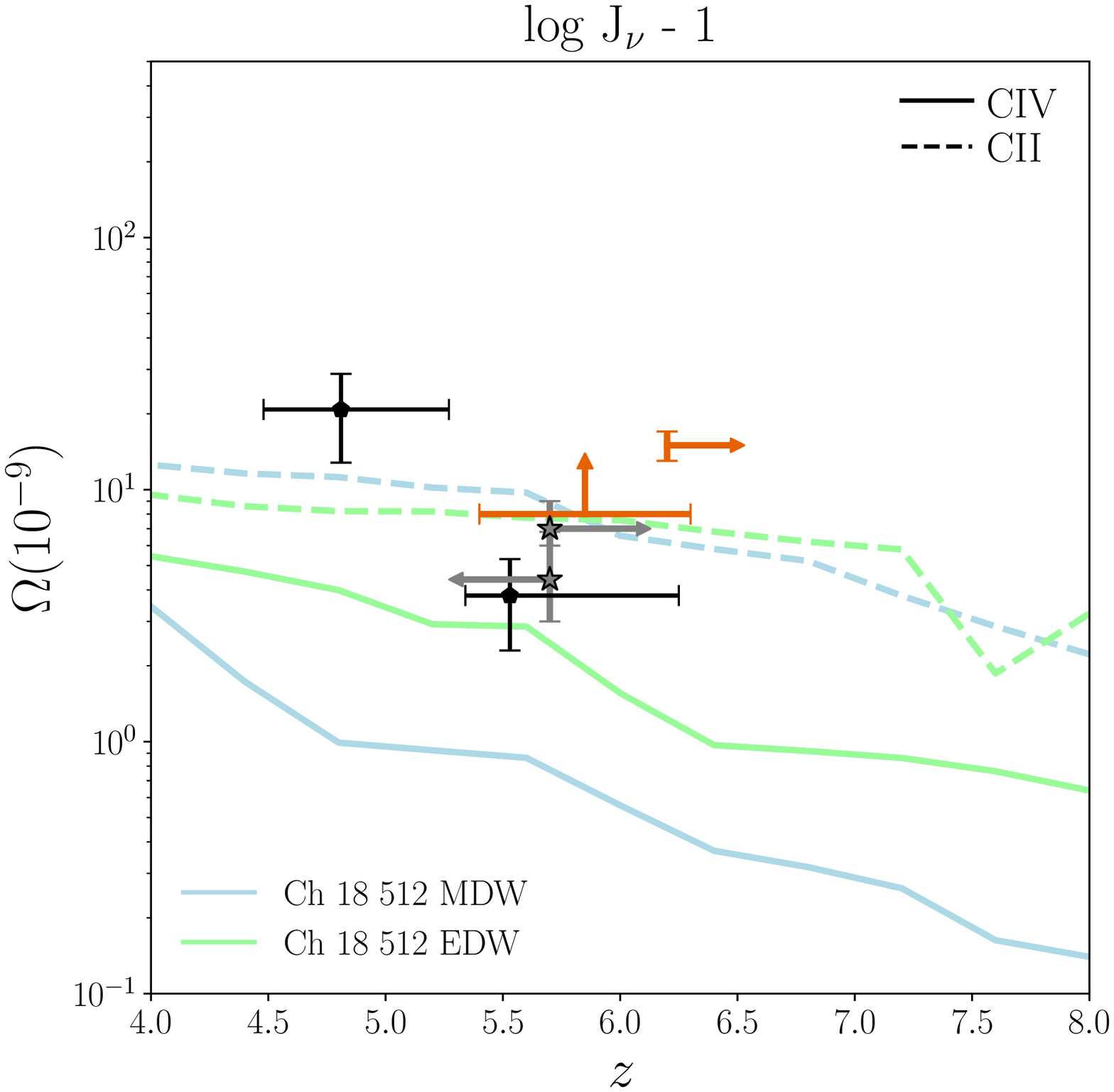}
        \caption{Evolution of the CII and CIV cosmological mass density when the normalization of the UVB is varied at 1 Ryd (comparison of MDW and EDW feedback prescriptions). On the left panel the results with the uniform {\small{HM12}} are presented. The right window panel shows results of the test \textbf{log J$_{\nu} -$ 1}. In both cases, the runs used are Ch 18 512 MDW and Ch 18 512 EDW. The solid lines mark the evolution of $\Omega_{\text{CIV}}$ for 13.8 $<$ $\log$ N$_{\text{CIV}}$(cm$^{-2}$) $<$ 15.0, and the dashed lines $\Omega_{\text{CII}}$ in the range 13.0 $<$ $\log$ N$_{\text{CII}}$(cm$^{-2}$) $<$ 15.0. The orange points with errors represent the observational lower limits for $\Omega_{\text{CII}}$ from \citet{becker2006}, and the grey arrows, the estimates from \citet{cooper2019}. There is not crossover of CII and CIV at any redshift, because of the different orders of magnitude of the mass densities of these ions. In addition, different feedback prescriptions do not seem to give rise to a remarkable distinction in the evolution of CII. Yet, the plot reveals that a softer UVB effectively favours low ionization states as CII, and brings down the gap between the observations from \citet{becker2006} and \citet{cooper2019} and the simulated column densities.}
\label{fig:omega_feed_both}
\end{figure}

Figure~\ref{fig:omega_nor} shows a comparison of the calculated CIV cosmological mass density at 4 $< z <$ 8 for synthetic absorbers in the range 13.8 $<$ (log N$_{\text{CIV}}$/ cm$^{-2}$) $<$ 15.0 and observations by \citet{pettini2003} and \citet{ryanweber2009} in orange circles, \citet{codoreanu2018} in yellow circles, \citet{songaila2001,songaila2005} in cyan triangles, \citet{meyer2019} in magenta triangles, \citet{simcoe2011} in dark green inverted triangles, \citet{dodorico2013} in pink squares, \citet{boksenberg2015} in grey diamond, upper limits from \citet{bosman2017} in purple star and \citet{diaz2016} in black pentagons.\newline
As a consequence of the discrepancy presented in Figures~\ref{fig:c4_cddf_48_5}, \ref{fig:c4_cddf_56_5} and \ref{fig:c4_cddf_64_5}, the CIV cosmological mass density in the right hand panel of Figure~\ref{fig:omega_nor} is at least an order of magnitude below the reference case on the left panel, because the number of absorbers in the range 13.8 $<$ $\log$ N$_{\text{CIV}}$(cm$^{-2}$) $<$ 15.0 are underproduced by the simulations in the framework of the modified UVB. The most remarkable difference is visible in the reference run Ch 18 512 MDW, with an order of magnitude shift between the blue curve (on the left) and the light blue one (on the right). \newline
Therefore, strong variations of the UVB around 1 Ryd (specifically, normalization changes in the wavelength range where the transition occurs) seem to have a large impact on the number of CIV absorbers at all redshifts. Fi\-gu\-res~\ref{fig:omega_mol_both} and \ref{fig:omega_feed_both} draw a comparison of the evolution of CII and CIV in the redshift range 4 $< z <$ 8, with the original {\small{HM12}} (on the left panel) and the test \textbf{log J$_{\nu} -$ 1} (on the right). As discussed above, the cosmological mass density of CIV is underrepresented. However, the re\-sul\-ting mass density of CII sig\-ni\-fi\-cant\-ly improves with a softer UVB, indicating that the hypothesis made to perform this test is well-motivated. Effectively, the number of large column density CII absorbers increases and the right panels of Figures~~\ref{fig:omega_mol_both} and \ref{fig:omega_feed_both} are now compatible with the limits measured by \citet{becker2006} and the most recent estimates made by \citet{cooper2019}.\newline
The feedback prescription does not play a major role in the evolution of CII in this redshift range, while the molecular cooling run Ch 18 512 MDW mol shows a relatively good agreement with the observational data.\newline
Due to the different orders of magnitude between the calculated mass densities of CII and CIV, the right panels of Figures~\ref{fig:omega_mol_both} and \ref{fig:omega_feed_both} show no crossover of the low and high ionization states of Carbon. A natural conclusion from this could be that decreasing the intensity of the UV background lead to an improvement in low ionization states at the expense of a poor calculation of CIV absorbers, that are traditionally well constrained by observations.\newline

As pointed out before, CIV is not well represented by this variation of the UVB, but there is a good improvement in the column densities of CII. However, it is difficult to draw definitive conclusions from the column density relationships because the number of absorbers depends strongly on the ion, with less CIV synthetic absorbers in the case of \textbf{log J$_{\nu} -$ 1}. We suggest variations in the UVB lower than the original HM12 emissivity, but not below to an order of magnitude to preserve the improvements in the low-ionization states while keeping positive results in CIV.

\section{Variation of the assumed HI self--shielding prescription}
\label{ssh}

A final test that can be done with our simulations is a variation of the HI self--shielding prescription. \citet{garcia2017b} briefly discuss the need for low ionization states self--shielding (SSh) treatment, however, here we try to quantify the impact of a HI self--shielding prescription different than the extensively used model by \citet{rahmati2013a} with the HM12 model. For this purpose, we use the new HI SSh model described by \citet{chardin2017b}. Their best fitting parameters as a function of redshift have been obtained with radiative transfer simulations (calibrated with Ly$\alpha$ forest data after the EoR). One of the merits of these models is that they focused on redshifts corresponding to Reionization. Instead, \citet{rahmati2013a} HI SSh prescription is valid up to $z =$ 5, which constitutes a limitation when applying this formulation in our models, as commented in \citet{garcia2017b}.  Here we compare results for the metal ions with HI SSh treatments by \citet{rahmati2013a} and \citet{chardin2017b} -hereafter R13 and C17, respectively-.\\

The evolution of the photoinization rate $\Gamma_{\text{phot}}$ is calculated using RT codes, such that:
\begin{equation}\label{eq:ssh}
\frac{\Gamma_{\text{phot}}}{\Gamma_{\text{UVB}}}=(1-f) \left[1+\left(\frac{n_{\text{H}}}{n_0} \right)^{\beta}\right]^{\alpha_1} + f  \left[ 1+\left(\frac{n_{\text{H}}}{n_0} \right)\right]^{\alpha_2},
\end{equation} 
\noindent where $\Gamma_{\text{UVB}}$ is the photoionization rate as a function of redshift and it is assumed from the UVB field. Here, $f$, $\alpha_1$, $\alpha_2$, $\beta$ and $n_0$ are free parameters of the model, calculated with RT and, the number density of Hydrogen $n_{\text{H}}$ and the temperature $T$ are taken directly from the numerical simulation used. \\

The best fitting parameters of the equation~\ref{eq:ssh} as a function of redshift found by R13 and C17 are shown in Table~\ref{table_ssh}.
\begin{table}[h!]
\caption{\footnotesize{Best fitting parameters for SSh modelling}}
\begin{center}
\label{table_ssh}
\vspace{0.2cm}
\resizebox{0.95\textwidth}{!}{%
\begin{tabular}{|c|c|c|c|c|c|c|} 
\hline
Model & $z$ &  $n_0$ & $\alpha_1$ & $\alpha_2$ & $\beta$ &  $f$ \\
 &  & (cm$^{-3}$) &  &  & &\\ \hline
\textbf{R13} & 1 - 5  &1.003 $\pm$ 0.005 $n_{\text{H,SSh}}$  & -2.28 $\pm$ 0.31 & -0.84 $\pm$ 0.11 & 1.64 $\pm$ 0.19 & 0.02 $\pm$ 0.0089\\ \hline
  &  {$4.0$} &  {$0.009346$}  & {$-0.950010$} & {$-1.503310$} & {$5.873852$} & {$0.015308$} \\
  &  {$5.0$} &  {$0.010379$}  & {$-1.294665$} & {$-1.602099$} & {$5.056422$} & {$0.024356$} \\
\textbf{C17}  &  {$6.0$} &  {$0.006955$}  & {$-0.941372$} & {$-1.507124$} & {$6.109438$} & {$0.028830$} \\
  & {$7.0$} &  {$0.002658$}  & {$-0.866887$} & {$-1.272957$} & {$7.077863$} & {$0.040894$} \\
  &  {$8.0$} &  {$0.003974$}  & {$-0.742237$} & {$-1.397100$} & {$7.119987$} & {$0.041213$} \\
\hline
\end{tabular}
}
\vspace{0.2cm}
\end{center}
\footnotesize{The free parameters correspond to equation~\eqref{eq:ssh} from RT calculations derived by \citet{rahmati2013a}, R13, and \citet{chardin2017b}, C17. The term $n_{\text{H,SSh}}$ corresponds to the self--shielding density threshold.}\\
\end{table}

Differently than the outcome for the metal ions studied in the previous section, one would expect that HI would be more sensitive to a variation of the self--shielding prescription adopted for this transition. In fact, works by \citet{rahmati2013a} and \citet{chardin2017b} self-consistently calculate the distribution of neutral Hydrogen with RT codes, based on the number density of Hydrogen. We compute the following observables for HI: the column density distribution function (CDDF) $f_{\text{HI}}$, the HI cosmological mass density $\Omega_{\text{HI}}$, and the mass density associated to DLA systems $\Omega_{\text{DLA}}$.\newline

\begin{figure}[h!]
\centering
	\includegraphics[width=0.48\linewidth]{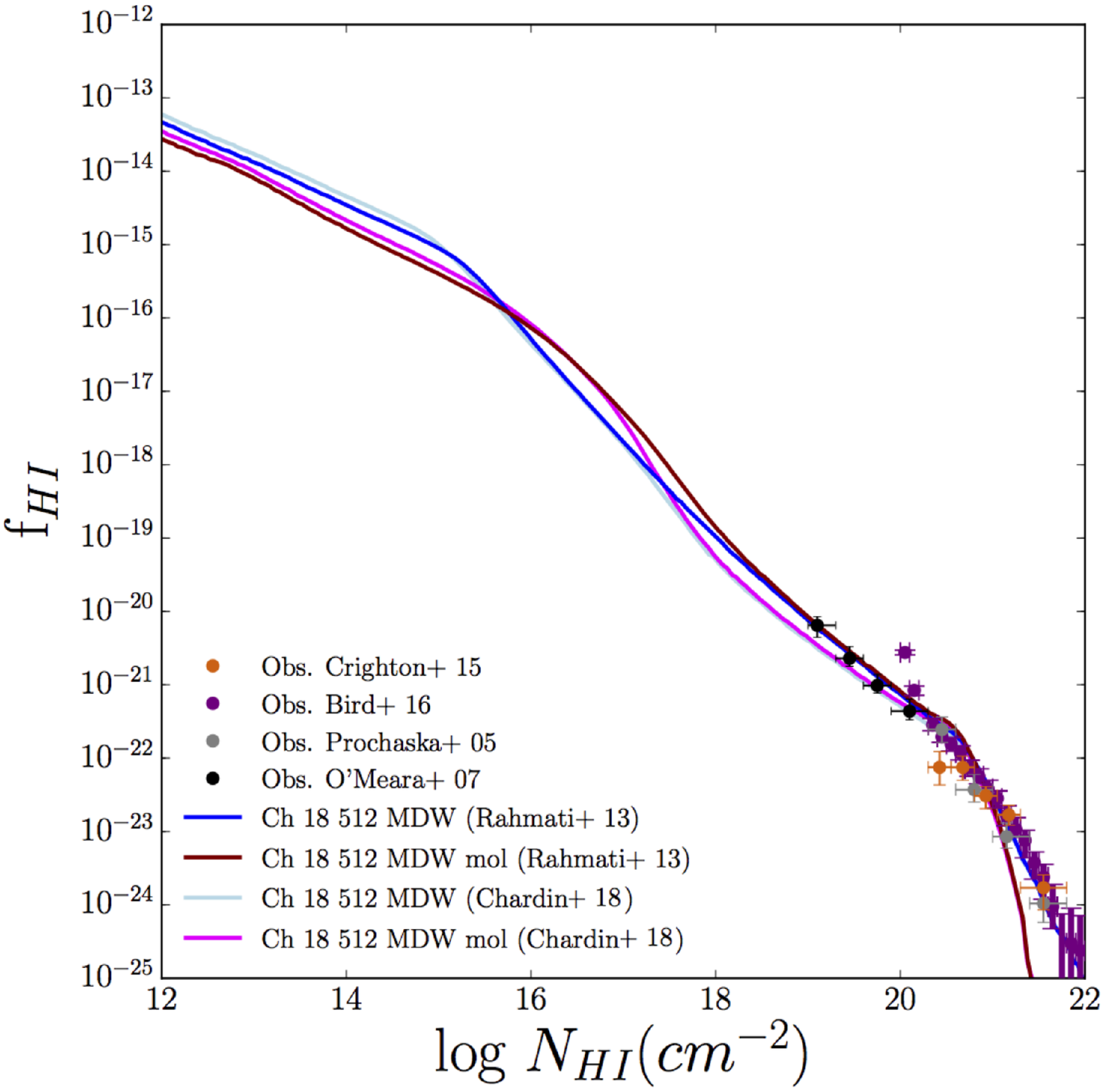}
	\includegraphics[width=0.48\linewidth]{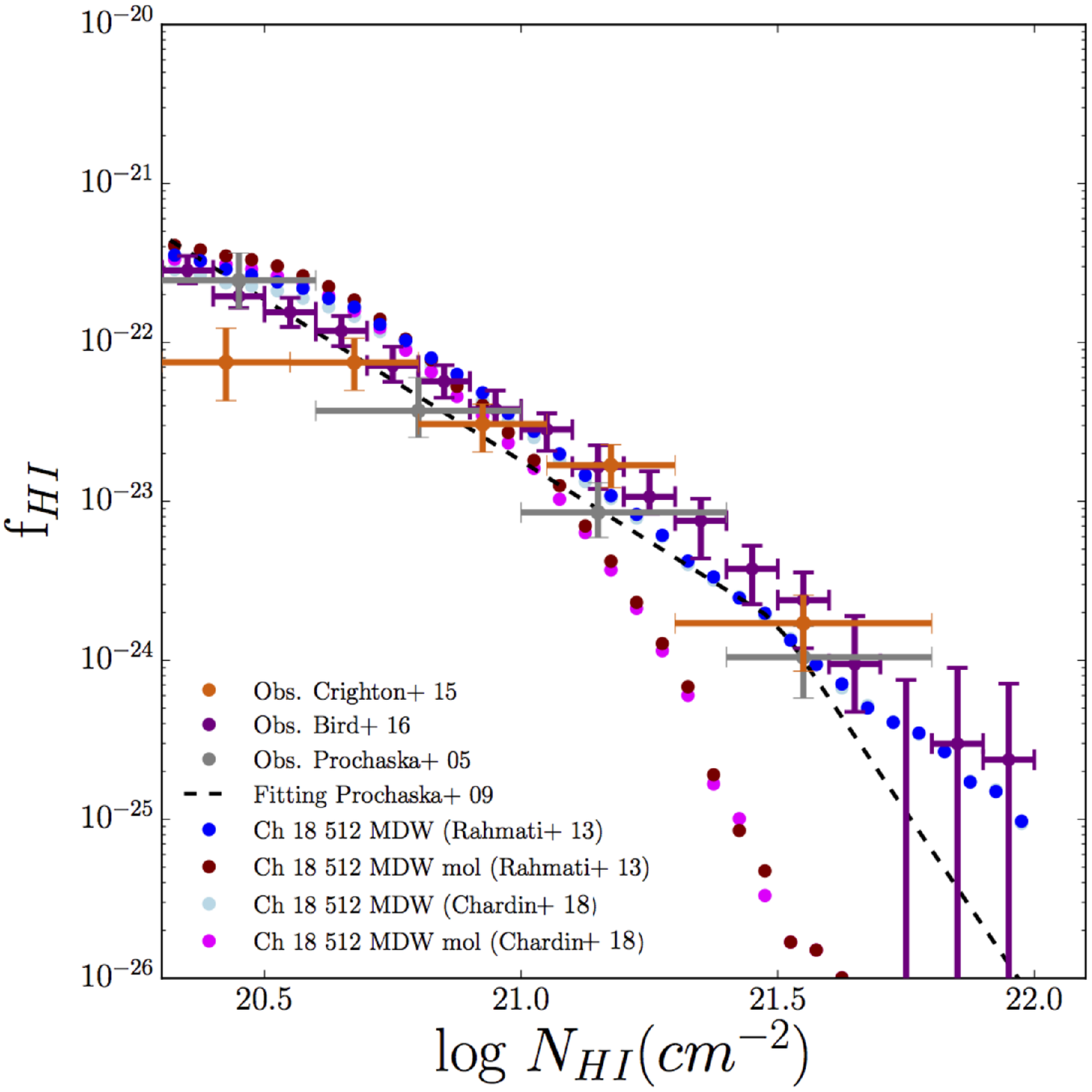}
        \caption{HI column density distribution function at $z =$ 4 comparing two self--shielding prescriptions \citep[][]{rahmati2013a,chardin2017b} in simulations with different molecular cooling content (Ch 18 512 MDW and Ch 18 512 MDW mol) and the HM12 model. On the left side is shown the simulated HI--CDDF in the range 12 $<$ $\log$ N$_{\text{HI}}$ (cm$^{-2}$) $<$ 22 and a comparison with observations by \citet{prochaska2005} in grey, \citet{omeara2007} in black, \citet{crighton2015} in orange and \citet{bird2016} in purple. In the right panel, the CDDFs are limited to the DLA regime (20.3 $<$ $\log$ N$_{\text{HI}}$ (cm$^{-2}$) $<$ 22) and compared with the fitting function by \citet{prochaska2009} for DLA systems at redshift 4.0--5.5 (black dashed line).}
        \label{fig:h1cddfs}
\end{figure}

In Figure~\ref{fig:h1cddfs} is shown the HI--CDDF at $z =$ 4 comparing the two self-shielding prescriptions (R13 and C17) and simulations with different molecular cooling contents, Ch 18 512 MDW and Ch 18 512 MDW mol. In addition, we compare our theoretical predictions with  observational detections of HI--CDDF at $z$ around 4 in two regimes: the range of column densities 12 $<$ $\log$ N$_{\text{HI}}$ (cm$^{-2}$) $<$ 22 in the left panel and a zoom around the DLAs region, 20.3 $<$ $\log$ N$_{\text{HI}}$ (cm$^{-2}$) $<$ 22, in the right panel. In the first case, observations by \citet{prochaska2005} in grey, \citet{omeara2007} in black, \citet{crighton2015} in orange and \citet{bird2016} in purple have been plotted. Instead, in the DLA zoom, a comparison has been drawn with the fitting function proposed by \citet{prochaska2009} for DLA systems at redshift 4.0--5.5.\newline

One can see a difference between the two self--shielding prescription implemented at $z =$ 4, \citet{chardin2017b} model predicts a larger amount of neutral Hydrogen hosted in systems in the Lyman--$\alpha$ forest and less systems in the sub-DLA and DLA regimes, indicating that the HI mass density should be larger with \citet{rahmati2013a} SSh prescription. In \citet{garcia2017b}, we find that the largest contribution to $\Omega_{\text{HI}}$ comes from systems with large column densities, in particular, DLAs. We predict higher values of $\Omega_{\text{HI}}$ with R13 HI SSh formulation than C17, regardless of the molecular cooling model considered. It is important to remember that at this redshift ($z =$ 4), both formulations are valid and their best fitting parameters are calibrated with observations, thus, our conclusions are not limited by different constraints of the HI SSh modelling.\newline
Interestingly, the number of systems in a column density bin at a given absorption path is barely affected by the SSh prescription, but it depends strongly on the chemistry of the molecules, and the gap between the number of systems is approximately fixed when comparing two simulations with and without molecules in Lyman alpha forest regime (blue and dark red, and light blue and magenta cases, respectively).\newline

\begin{figure}[h!]
\centering
	\includegraphics[width=0.48\linewidth]{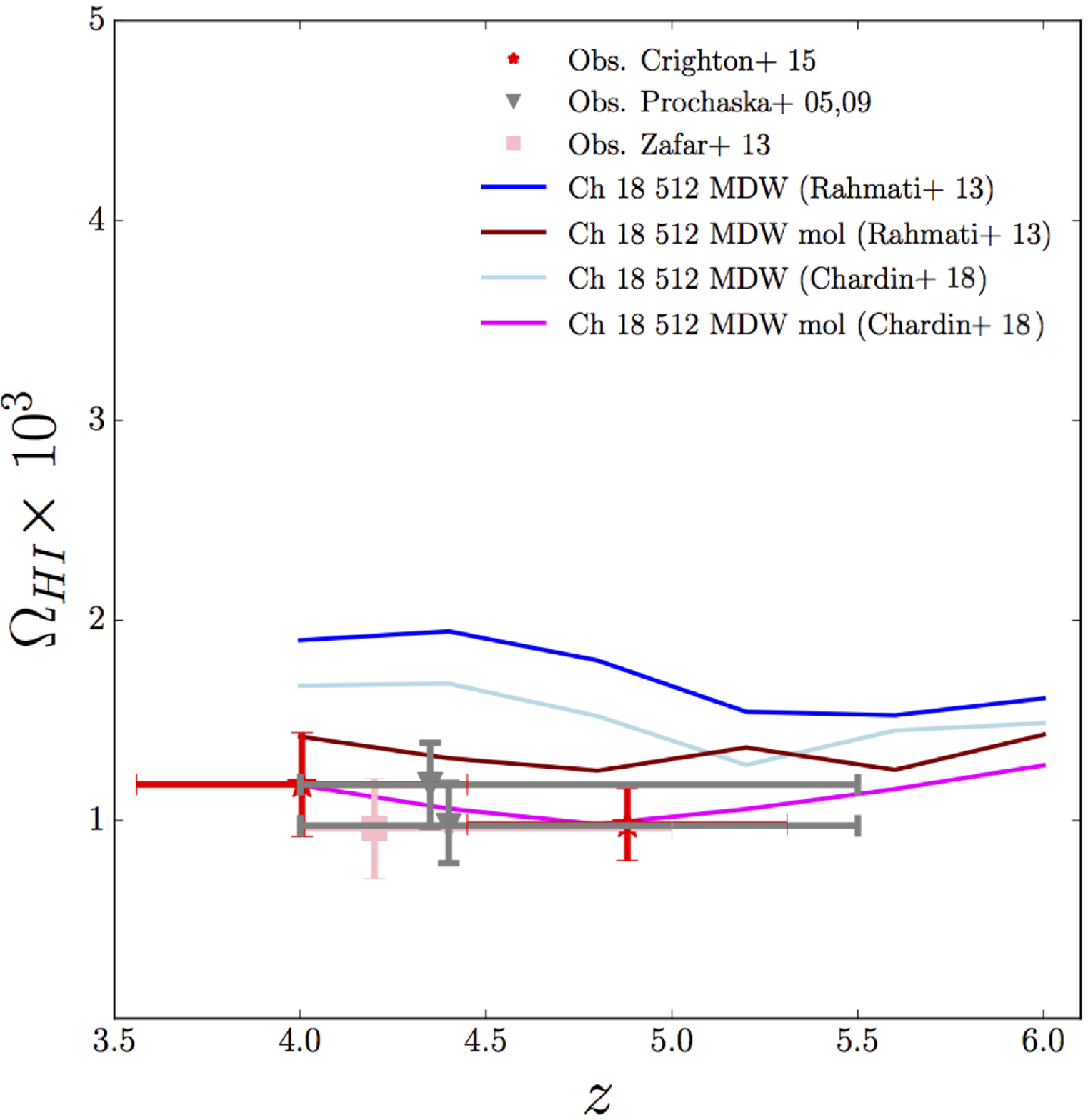}
	\includegraphics[width=0.48\linewidth]{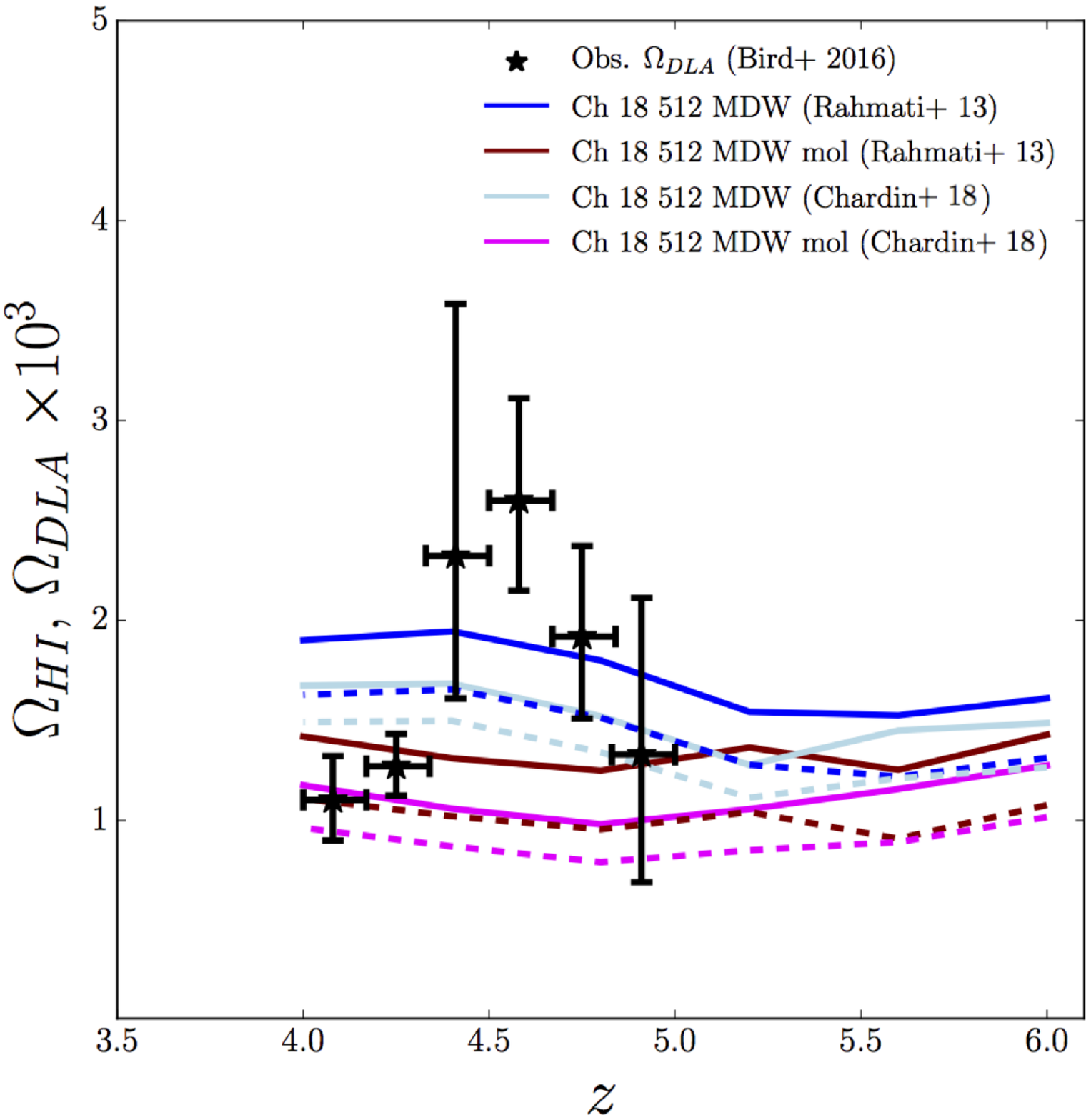}
        \caption{Cosmological mass density of HI with different HI self--shielding treatments: \citet{rahmati2013a} and \citet{chardin2017b}. The left diagram shows the prediction of the HI mass density from simulations with specific molecular cooling content (Ch 18 512 MDW and Ch 18 512 MDW mol) and a comparison with \citet{prochaska2005, prochaska2009} in grey inverted triangles, \citet{zafar2013} in pink square and \citet{crighton2015} in red stars. The right panel displays the cosmological mass density associated to neutral Hydrogen and DLAs, $\Omega_{\text{HI}}$ and $\Omega_{\text{DLA}}$ (solid and dashed lines, respectively) and the observational data for $\Omega_{\text{DLA}}$ by \citet{bird2016} in black stars.}
\label{fig:h1omegas}
\end{figure}

Figure~\ref{fig:h1omegas} (left panel) displays the cosmic mass density of neutral Hydrogen with the self--shielding prescriptions by \citealt{rahmati2013a} (blue and dark red for the runs Ch 18 512 MDW and Ch 18 512 MDW mol, respectively) and \citealt{chardin2017b} (light blue and magenta corresponding to Ch 18 512 MDW and Ch 18 512 MDW mol, respectively) and compares with observations by \citet{prochaska2005} and \citet{prochaska2009} in grey inverted triangles, \citet{zafar2013} in pink square and \citet{crighton2015} in red stars. As predicted above, the SSh prescription by \citet{rahmati2013a} gives rise to a larger amount of neutral Hydrogen when compared with the results for $\Omega_{\text{HI}}$ with \citet{chardin2017b} model. Interestingly, the introduction of this new self--shielding treatment reduces the tension between our models without molecular cooling and the observations at $z =$ 4 and accurately predicts the HI mass density when molecular cooling is taken into account.\newline
This is by far the most important effect of the implementation of a different self--shielding prescription in our simulations: the amount of HI mass density at redshifts between 4-6 is in better agreement with observational detections.\newline
Finally, we point out to the reader that models with molecular cooling included gives a better prediction of $\Omega_{\text{HI}}$ compared with observations, because they take into account the conversion of atomic to molecular Hydrogen at very high densities, where the self--shielding of HI is occurring.\newline

On the right hand side of Figure~\ref{fig:h1omegas} we show a comparison of $\Omega_{\text{HI}}$ and $\Omega_{\text{DLA}}$ (solid and dashed lines, respectively) with observations for $\Omega_{\text{DLA}}$ by \citet{bird2016}. The amount of HI mass density hosted in DLAs (dashed lines) converges at high redshift ($z \sim$ 6) for models with different SSh prescription when the molecular cooling content is fixed. This is indeed quite an interesting result because R13 SSh prescription has not been calibrated at this redshift, although it is extensively used in the literature at redshifts higher than 5. One can say that the use of the model is justified at $z =$ 6.\newline 
Additionally, there is an agreement between the predicted trends for $\Omega_{\text{DLA}}$ (with different self--shielding prescriptions and molecular cooling content models) and observations of this quantity by \citet{bird2016}.\newline

It is worth noting that the column density distribution functions in Figure~\ref{fig:h1cddfs} shows that molecular cooling is driven by the conversion of neutral Hydrogen to H2, that becomes important in high-density regions where new stars are being formed. The molecules cool the surrounding gas and the amount of atomic Hydrogen is less abundant. This effect is particularly important in the regime of DLAs \citep{garcia2017b}. What results more interesting is that molecular chemistry plays a more relevant effect in the amount of neutral Hydrogen than the HI self-shielding model itself, although the evolution of \citet{chardin2017b} model is much more dynamic at high redshift than \citet{rahmati2013a}'s one.\newline
Figure~\ref{fig:h1omegas} leads to the same conclusion in a different way: simulations without molecular cooling implemented (Ch 18 512 MDW) boost the amount of HI compared with Ch 18 512 MDW mol, regardless of the SSh prescription considered. Then, the cosmological mass density increases by a factor of 2 compared with simulations contained molecular cooling. The effect is more pronounced by contrasting HI and DLA cosmic mass density: there are two processes involved, the molecular cooling that converts HI into H2 and reduces the amount of HI in the DLA column density range, but also, the evolution of the HI-SSh modelling, that becomes more critical at high redshift (\citet{rahmati2013a} SSh stops being valid at $z =$ 6, whilst \citet{chardin2017b} model is compelling up to $z =$ 10).

\section{Discussion and Conclusions}\label{conclusions}
This work contains a compilation of variations to the uniform {\small{HM12}} ionizing field assumed in the photoionization models to obtained the results presented in \citet{garcia2017a} and \citet{garcia2017b}. We find that the large uncertainties on the UVB, especially at high redshift, require better cons\-traints for several probes, and of course, other improved UVB models.\newline

In a similar vein, observational detections of metal ionic species could be used to constrain the UVB spectral normalization in the wavelength range where these states occur (100-1000 \AA). A careful fine-tuning is required, but most of the steps are clear: a reduction in the normalization of the UV emissivity improves the number of absorbers with low ionization potential energy.\newline

When the emissivity of the assumed UVB is reduced by one dex, we find that the total number of CIV absorbers significantly decreases compared with the original HM12 and absorbers in the range of the detections by D'Odorico et al. (2013) are definitely not well represented (e.g. there is a discrepancy of the CIV comoving mass density and CIV CDDF with the observations). Instead, there is a moderate improvement in the calculated number of absorbers for SiIV with a softer UVB, in agreement with findings by \citet{bolton2011}.\newline

It is particularly encouraging that the number of absorbers of CII, SiII and OI in the range of the observations also rises when the UVB implemented is softer, independently of the simulation used, and therefore, the estimate of the column densities of these low ionization states improves, justifying the hypothesis that led us to do this test. Additional work can be done in this direction, with a more moderated reduction in the emissivity of the UVB that would give better predictions for the column density of low ionization states, and be more compatible with the CIV absorbers incidence rate inferred with \citet{dodorico2013} observations and our mock spectra for that ion.\newline

We have also computed some Hydrogen statistics up to $z =$ 6. We have limited our calculations to this redshift for two reasons: 1) at high redshift a uniform UVB does not provide a good representation of the rapid evolution of HI during Reionization; 2) the self--shielding prescription from \citet{rahmati2013a} is calibrated with the photoionization rates of {\small{HM12}}. To alleviate the absence of a self-consistent SSh at higher redshifts, we have compared our previous results with a recent SSh  prescription by \citet{chardin2017b}, whose best-fit parameters have been calibrated up to $z =$ 10.\newline
We find that at $z =$ 4, the HI CDDF predicted by \citet{chardin2017b} model produces more neutral Hydrogen in systems with low column densities (the Lyman--$\alpha$ forest) and less systems in  the sub-DLAs and DLAs range. This result leads to a higher mass densities of HI, $\Omega_{\text{HI}}$, with \citet{rahmati2013a} SSh prescription than \citet{chardin2017b}, independent on the molecular cooling model considered. At $z =$ 4, both formulations \citep{rahmati2013a, chardin2017b} have been calibrated with observations, therefore, our results for the HI CDDF and the cosmological mass density of neutral Hydrogen are valid, regardless of the assumptions of each HI self--shielding model.\newline

Regardless of the intensity of the UVB field considered, we found a non-negligible number of ion absorbers with column densities above log N $\geq$ 15 cm$^{-2}$. This prediction from our models can be tested with future observations that in principle will be able to detect rarer high column density systems. In addition, an appreciable difference in the number of high column density absorbers would be detected if we trace high gas overdensity, but our pipeline is currently not tracing gas in this region for statistical reasons, since we produce random sightlines inside the box.\newline

As discussed in \citet{garcia2017b}, we confirm that different feedback prescriptions (EDW, MDW) have not made a significant difference in the calculation of the number of absorbers of the ionic transitions. The same conclusion is also true for different models with/without molecular cooling for observables associated to the metal ions. Instead, the low temperature and molecular cooling module has proven to be quite important to match the calculated HI CDDF, $\Omega_{\text{HI}}$ and $\Omega_{\text{DLA}}$ with observations available at $z =$ 4 - 6. \newline

The numerical runs and the pipeline of this paper have been calibrated to resemble the physical conditions of the gas residing in the CGM and IGM at $z \sim$ 6. Most of the observables at that time of the Universe are well represented by our theoretical models. Nonetheless, there are a few caveats in this work: first, more observations of metal absorption lines in the spectra of high redshift quasars are required. Although, we have included the largest catalogue of HI and ionic detections at high redshift, more observational data is needed to better constrain the emissivity of the UVB. Second, the numerical resolution is still not enough to fully trace the environments where the low ionization states lie and simultaneously, describe the IGM. New generation of numerical simulations will improve the description of these species, without sacrificing the physics occurring in the inter- and circumgalactic medium. Finally, we stress that radiative transfer effects are not included in our models, and therefore, we do not follow the progression of Reionization, nor the evolution of the HII bubbles or their topology. The implicit assumption is that our boxes (that are small compared to the size of the HII bubbles at the redshifts of interest) represent a region of the Universe already reionized at a level given by the HM12 UVB. At 6 $< z <$ 8, chemical enrichment occurs mostly inside and in close proximity of galaxies (interstellar medium, CGM and high density IGM) where, assuming an inside-out progression of Reionization, the gas in which metals lie should be ionized. Although proper RT calculations would be more accurate, they are extremely expensive from the computational point of view and they could be done in the future.\newline

A fundamental issue persists in the field from the numerical point of view: it is extremely challenging to model the low ionization states present in the gas and provide a good description of the environment where they lie, mainly due to insufficient resolution and a proper self-shielding treatment for the ions. We have considered only the effect of HI self-shielding \citep{rahmati2013a,chardin2017b}, but did not introduce any self-shielding of low ionization absorbers (which lie in clumpy structures). A first attempt has been proposed for DLA systems at low redshift by \citet{bird2016}. However, current works miss this component at high redshift because it is still not well understood how high density regions self-shield the gas during the progression of Reionization.\newline

In summary, we have compared HI and ion observables available at high redshift and found that most of the results discussed are compatible in the redshift range 4 $< z <$ 8. When discrepancies between observations and synthetic calculations have arised, we provide a physical explanation of the nature of such differences. It is worth noting that all results from mock spectra will be improved in the future with more observational detections of ion absorbers in the high redshift quasar spectra. \newline


\begin{thebibliography}
\bibitem[Abel et al.(2002)]{abel2002} Abel, T., et al.\ 2002, Science, 295, 93
\bibitem[Planck Collaboration (2015)]{planck2015} Ade, P.~A.~R., et al.\ 2015, \aap, 594, A13 
\bibitem[Asplund et al.(2009)]{asplund2009} Asplund, M. et al.\ 2009,  \araa, 47, 481
\bibitem[Barnes et al.(2009)]{barnes2009} Barnes, L.~A. \& Haehnelt, M.~G.\ 2009, \mnras, 397, 511  
\bibitem[Becker et al.(2006)]{becker2006}  Becker, G.~D et al.\ 2006, \apj, 640, 69  
\bibitem[Becker et al.(2011)]{becker2011}  Becker, G.~D et al.\ 2011, \apj, 735, 93 
\bibitem[Becker et al.(2019)]{becker2019}  Becker, G.~D. , et al.\ 2019, \apj, 883, 163
\bibitem[Bird et al.(2014)]{bird2014} Bird, S., et al.\ 2014, \mnras, 445, 2313
\bibitem[Bird et al.(2015)]{bird2015} Bird, S., et al.\ 2015, \mnras, 447, 1834  
\bibitem[Bird et al.(2016)]{bird2016} Bird, S. ,, et al.\ 2016 \mnras, 466, 2111
\bibitem[Boksenberg et al.(2015)]{boksenberg2015}  Boksenberg, A. and {Sargent}, W.~L.~W.\ 2015, \apjs, 218, 7
\bibitem[Bolton et al.(2005)]{bolton2005}  Bolton, J.~S., et al.\ 2005, \mnras, 357, 1178
\bibitem[Bolton et al.(2011)]{bolton2011}  Bolton, J.~S. \& Viel, M.\ 2011, \mnras, 414, 241   
\bibitem[Bosman et al.(2017)]{bosman2017} Bosman, S.~E.~I., et al.\ 2017, \mnras, 470, 1919
\bibitem[Bromm et al.(2002)]{bromm2002} Bromm, V., et al.\ 2002, \apj, 564, 23  
\bibitem[Cai et al.(2017)]{cai2017} Cai, Z., et al.\ 2017, \apjl, 849, L18
\bibitem[Carswell et al.(2014)]{vpfit} Carswell, R.~F. \& Webb, J.~K.\ 2014, (astro-ph/1408.015)
\bibitem[Cen et al.(2011)]{cen2011}  Cen, R. and Chisari, N.~E.\ 2011, \apj, 731, 11
\bibitem[Chabrier (2003)]{chabrier2003} Chabrier, G. \ 2003, \pasp, 115, 763
\bibitem[Chardin et al.(2017b)]{chardin2017b} Chardin, J.  , et al.\ 2017, \mnras, 478, 1065
\bibitem[Ciardi et al.(2005)]{ciardi2005} Ciardi, B. \& Ferrara, A.\ 2005, \ssr, 116, 625  
\bibitem[Codoreanu et al.(2018)]{codoreanu2018} Codoreanu , et al.\ 2018, \mnras, 481, 4940  
\bibitem[Cooke et al.(2014)]{cooke2014} Cooke, J., et al.\ 2014, \mnras, 441, 837
\bibitem[Cooper et al.(2019)]{cooper2019}  Cooper, T.~J. , et al.\ 2019,  \apj, 882, 77
\bibitem[Crighton et al.(2015)]{crighton2015} Crighton, N.~H.~M., et al.\ 2015, \mnras, 452, 217
\bibitem[D\'iaz et al.(2016)]{diaz2016} D\'iaz, C.~G., et al.\ 2016, Bolet\'in de la Asociaci\'on Argentina de Astronom\'ia, 58, p.51-53 
\bibitem[Dijkstra et al.(2004)]{dijkstra2004} Dijkstra, M., et al.\ 2004, \apj, 613, 646
\bibitem[D'Odorico et al.(2013)]{dodorico2013} D'Odorico, V., et al.\ 2013, \mnras, 435, 1198  
\bibitem[Doughty et al.(2018)]{doughty2018} Doughty, C., et al.\ 2018, \mnras, 475, 4717  
\bibitem[Doughty et al.(2019)]{doughty2019} Doughty, C., et al.\ 2019, \mnras, 489, 2755
\bibitem[Ferland et al.(2013)]{ferland2013} Ferland, G.~J., et al.\ 2013, Revista Mexicana de Astronom\'ia y Astrof\'isica, 49, pp. 137-163
\bibitem[Finlator et al.(2013)]{finlator2013} Finlator, K., et al.\ 2013, \mnras, 436, 1818
\bibitem[Finlator et al.(2015)]{finlator2015} Finlator, K., et al.\ 2015, \mnras, 447, 2526
\bibitem[Finlator et al.(2016)]{finlator2016} Finlator, K., et al.\ 2016, \mnras, 459, 2299
\bibitem[Finlator et al.(2018)]{finlator2018} Finlator, K. , et al.\ 2018, \mnras, 480, 2628 
\bibitem[Garc\'ia et al.(2017a)]{garcia2017a} Garc{\'i}a, L.~A., et al.\ 2017a, \mnras, 469, L53
\bibitem[Garc\'ia et al.(2017b)]{garcia2017b} Garc{\'i}a, L.~A.  et al.\ 2017b, \mnras, 470, 2494
\bibitem[Gnat \& Ferland (2012)]{gnat2012} Gnat O., Ferland G. J., \ 2012, \apjs, 199, 20
\bibitem[Haardt et al.(1999)]{haardt1999}  Haardt, F.\ 1999, \memsai, 70, 261
\bibitem[Haardt et al.(2001)]{haardt2001}  Haardt, F. \& Madau, P.\ 2001, (XXXVI Rencotres de Moriond, 2001)
\bibitem[Haardt et al.(2012)]{haardt2012}  Haardt, F. \& Madau, P.\ 2012, \apj, 746, 125
\bibitem[Haiman et al.(2015)]{haiman2015} Haiman, Z.\ 2015, Astrophysics and Space Science Library, 423, 1 
\bibitem[Hassan et al.(2017)]{hassan2017} Hassan, S., et al.\ 2017, \mnras, 473, 227 
\bibitem[Katsianis et al. (2015)]{katsianis2015} Katsianis, A., et al. \ 2015, \mnras, 448, 3001
\bibitem[Katsianis et al. (2016)]{katsianis2016} Katsianis, A., et al. \ 2016, Publications of the Astronomical Society of Australia, 33, 16 
\bibitem[Katsianis et al. (2017)]{katsianis2017} Katsianis, A., et al. \ 2017, \mnras, 464, 4977
\bibitem[Keating et al.(2014)]{keating2014} Keating, L.~C., et al.\ 2014, \mnras, 438, 1820
\bibitem[Keating et al.(2016)]{keating2016} Keating, L.~C., et al.\ 2016, \mnras, 461, 606
\bibitem[Kennicutt(1998)]{kennicutt1998} Kennicutt, Jr., R.~C.\ 1998, \apj, 498, 541
\bibitem[Lidz et al.(2006)]{lidz2006} Lidz, A., et al.\ 2006, \apjl, 639, L47  
\bibitem[Lidz et al.(2016)]{lidz2016} Lidz, A.\ 2016, (astro-ph/1511.01188)
\bibitem[Maio et al.(2007)]{maio2007} Maio, U., et al.\ 2007, \mnras, 379, 963
\bibitem[Maio et al.(2015)]{maio2015} Maio, U \& Tescari, E.\ 2015, \mnras, 453, 3798    
\bibitem[Mellema et al.(2006)]{mellema2006} Mellema, G., et al.\ 2006, \mnras, 372, 679 
\bibitem[Meyer et al.(2019)]{meyer2019} Meyer, R.~A. , et al.\ 2019, \mnras,  483, 19
\bibitem[Mortlock et al.(2011)]{mortlock2011} Mortlock, D.~J., et al.\ 2011, \nat, 474, 616  
\bibitem[Nagamine et al.(2004)]{nagamine2004} Nagamine, K., et al.\ 2004, \mnras, 348, 421
\bibitem[O'meara et al.(2007)]{omeara2007} O'Meara, J.~M., et al.\ 2007, \apj, 656, 666
\bibitem[Oppenheimer et al.(2006)]{oppenheimer2006} Oppenheimer, B.~D. \& Dav\'e, R.\ 2006, \mnras, 373, 1265
\bibitem[Oppenheimer et al.(2009)]{oppenheimer2009} Oppenheimer, B.~D. , et al.\ 2009, \mnras, 396, 729
\bibitem[Oppenheimer et al.(2013)]{oppenheimer2013} Oppenheimer, B.~D. \& Schaye, J.\ 2013, \mnras, 434, 1043
\bibitem[Padovani \& Matteucci (1997)]{padovani1993} Padovani, P., Matteucci, \ 1993, \apj, 416, 26
\bibitem[Pallottini et al.(2014)]{pallottini2014} Pallottini, A., et al.\ 2014,  \mnras, 440, 2498
\bibitem[Pettini et al.(2003)]{pettini2003} Pettini, M., et al.\ 2003, \apj, 594, 695
\bibitem[Pontzen et al.(2008)]{pontzen2008}  Pontzen, A., et al.\ 2008, \mnras, 390, 1349
\bibitem[Prochaska et al.(2005)]{prochaska2005} Prochaska, J.~X., et al.\ 2005,\apj, 635,123 
\bibitem[Prochaska et al.(2009)]{prochaska2009} Prochaska, J.~X., et al.\ 2009,\apj, 696, 1543
\bibitem[Puchwein et al.(2013)]{puchwein2013} Puchwein, E.  \& Springel, V.\ 2013, \mnras, 428, 2966
\bibitem[Rahmati et al.(2013a)]{rahmati2013a} Rahmati, A., et al.\ 2013, \mnras, 430, 2427  
\bibitem[Rahmati et al.(2013b)]{rahmati2013b} Rahmati, A., et al.\ 2013, \mnras, 431, 2261
\bibitem[Rahmati et al.(2015)]{rahmati2015} Rahmati, A., et al.\ 2015, \mnras, 452, 2034
\bibitem[Rahmati et al.(2016)]{rahmati2016} Rahmati, A., et al.\ 2016, \mnras, 453, 310
\bibitem[Ryan-Weber et al.(2009)]{ryanweber2009}  Ryan-Weber, E.~V., et al.\ 2009, \mnras, 395, 1476
\bibitem[Scannapieco et al.(2006)]{scannapieco2006} Scannapieco, E., et al.\ 2006, \mnras, 365, 615  
\bibitem[Simcoe (2006)]{simcoe2006} Simcoe, R.~A.\ 2006, \apj, 653, 977  
\bibitem[Simcoe et al.(2011)]{simcoe2011} Simcoe, R.~A., et al.\ 2011, \apj, 743, 21
\bibitem[Smith et al.(2017)]{smith2017} Smith, A., et al.\ 2017, \mnras, 464, 2963  
\bibitem[Songaila et al.(2001)]{songaila2001} Songaila, A.\ 2001, \apjl, 561, L153  
\bibitem[Songaila et al.(2005)]{songaila2005} Songaila, A.\ 2005, \aj, 130, 1996
\bibitem[Springel \& Hernquist (2003)]{springel2003} Springel, V., Hernquist, L.\ 2003, \mnras, 339, 289
\bibitem[Springel (2005)]{springel2005c}  Springel, V., \ 2005, \mnras, 364, 1105
\bibitem[Tescari et al.(2009)]{tescari2009} Tescari, E., et al.\ 2009, \mnras, 397, 411
\bibitem[Tescari et al.(2011)]{tescari2011} Tescari, E., et al.\ 2011, \mnras,  411, 826
\bibitem[Tescari et al.(2014)]{tescari2014} Tescari, E., et al.\ 2014, \mnras, 438, 3290
\bibitem[Thielemann et al. (2003)]{thielemann2003} Thielemann F.-K. et al. \ 2003, Nucl. Phys. A, 718, 139
\bibitem[Tornatore et al.(2007)]{tornatore2007} Tornatore, L., et al.\ 2007, \mnras, 382, 1050 
\bibitem[van den Hoek \& Groenewegen (1997)]{vandenhoek1997} van den Hoek, L. B., Groenewegen M. A. T., \ 1997, \aaps, 123, 305 
\bibitem[Wolfe et al.(2005)]{wolfe2005} Wolfe, A.~M., et al.\ 2005, \araa, 43, 861  
\bibitem[Woosley \& Weaver (1997)]{woosley1995} Woosley S. E., Weaver T. A., \ 1995, \apjs, 101, 181
\bibitem[Yoshida et al.(2003)]{yoshida2003} Yoshida, N., et al.\ 2003, \apj, 598, 73
\bibitem[Zafar et al.(2013)]{zafar2013} Zafar, T., et al.\ 2013, \aap, 556, A141
\bibitem[Zaroubi et al.(2007)]{zaroubi2007}  Zaroubi, S., et al.\ 2007, \mnras, 375, 1269
\end{thebibliography}
\end{document}